\documentclass[nofootinbib,superscriptaddress]{revtex4}

\usepackage{amsmath,amssymb,url}
\usepackage{rotating}
\usepackage{feynmf}
\usepackage[usenames]{color}
\usepackage{hyperref}

\def\eq#1{(\ref{#1})}

\def\eg{{\it e.g.\;}}

\def\gev{~{\ensuremath\rm GeV}}
\def\tev{\text{TeV}}

\def\mkk{m}
\def\mstar{M_\star}
\def\mpl{M_\text{Pl}}
\def\s{\mathcal{S}}
\def\sm#1#2{\mbox{\small{$ \frac{#1}{#2} $}}}
\def\sd{S_{n-1}}

\def\mtrans{\Lambda_T}
\def\deltakk{\delta_\text{KK}}	

\begin{document}

\title{Asymptotic safety and  Kaluza-Klein gravitons  at the LHC}

\author{Erik Gerwick}
\email{e.gerwick@sms.ed.ac.uk}
\affiliation{SUPA, School of Physics \& Astronomy, 
             University of Edinburgh, UK}

\author{Daniel Litim}
\email{d.litim@sussex.ac.uk}
\affiliation{Department of Physics \& Astronomy, 
               University of Sussex, Brighton BN1 9QH, UK}

\author{Tilman Plehn}
\email{plehn@uni-heidelberg.de}
\affiliation{Institut f\"ur Theoretische Physik, 
             Universit\"at Heidelberg, Germany}

\begin{abstract}
 We study Drell-Yan production at the LHC in 
low-scale quantum gravity models with extra dimensions.  Asymptotic safety implies 
that the ultra-violet behavior of gravity  
is dictated by a  fixed point. We show how the energy dependence 
of Newton's coupling regularizes the gravitational amplitude 
using a renormalization group improvement. 
We study LHC predictions
and find that Kaluza-Klein graviton signals are 
well above Standard Model backgrounds. This leaves a significant 
sensitivity to the energy scale $\Lambda_T$ where the gravitational 
couplings cross over from classical to fixed point scaling.   
\end{abstract}
\maketitle

\begin{fmffile}{SM}

\section{Introduction}
\label{sec:intro}

By allowing the fundamental Planck scale to be as low as the electroweak scale, models with extra-dimensions have raised enormous interest in the past decade. 
This way, the notorious hierarchy problem of the Standard Model (SM) is elegantly circumvented \cite{add,randall_sundrum}. In these scenarios, SM particles propagate on the four-dimensional brane, whereas gravity lives in the higher-dimensional bulk.  If realized in Nature, high-energy particle collisions at the LHC become sensitive to Planck scale physics and offer the exciting possibility to experimentally test the quantization of gravity. \medskip 

A description of graviton mediated scattering processes at energies close to the Planck scale requires a fundamental quantum theory for gravity. Unfortunately, perturbative quantization of four- and higher-dimensional gravity seems not applicable, and a fully satisfying theory is presently not at hand. In recent years, field-theoretical studies indicate that higher-dimensional gravity becomes asymptotically safe  \cite{Litim:2003vp,Fischer:2006fz}, meaning that the ultra-violet (UV) behavior of gravity is dictated by a non-perturbative fixed point under the renormalization group (RG) \cite{weinberg}. In the asymptotic safety scenario for gravity, the energy-dependence of the gravitational couplings is characterized by an energy scale $\Lambda_T$ where the gravitational couplings cross over from classical behavior at low energies to fixed point scaling at high energies \cite{Litim:2006dx,Niedermaier:2006ns,Percacci:2007sz,Litim:2008tt}. The non-trivial UV fixed point of gravity can be seen as the gravitational analogue of the perturbative UV fixed point of Yang-Mills theory. In gravity, the onset of fixed point scaling manifests itself through a weakening of Newton's coupling. Experimental signatures for asymptotic safety within low-scale quantum gravity are discussed in~\cite{Litim:2003vp,Fischer:2006fz,lp,tom_joanne,Falls:2010he}.  \medskip

A generic prediction of large extra dimensions is  
a tower of massive Kaluza-Klein gravitons. At colliders, Kaluza-Klein 
gravitons appear both as real final states \cite{lhc_real} and virtual intermediate 
states~\cite{limits_tevatron,lhc_ed_general}.  Real gravitons leave the 
detector unnoticed and appear as missing energy. 
Virtual gravitons mediate Drell-Yan processes \cite{dy}, 
and lead to deviations in Standard Model reference processes such as $pp \to \ell^+\ell^-$.
Missing energy signals and the study of benchmark spectra  are  standard search tools 
for physics beyond the SM. \medskip

Previously, gravitational Drell-Yan production was studied within  asymptotically safe gravity~\cite{lp,tom_joanne}, string theory~\cite{lhc_strings} and effective theory~\cite{grw,gps} (see \cite{uvcomp_review} for a review).  The relevant amplitude involves a sum over all intermediate Kaluza-Klein gravitons. Within effective theory, this sum is strongly UV divergent in two and more extra dimensions and requires a UV cutoff  of the order of the fundamental Planck scale. This scale cannot be precisely determined by low-energy considerations alone. Within asymptotically safe gravity, high energy interactions become dynamically suppressed \cite{Fischer:2006fz}, rendering the single graviton amplitude finite~\cite{lp}.  The amplitude no longer requires an artificial UV cutoff and is solely determined through the dynamical scale $\Lambda_T$.  Form-factor approximations  of the fixed point which treat the Kaluza-Klein gravitons perturbatively also require a UV cutoff, which ideally should be fixed by quantum gravity itself \cite{tom_joanne}.\medskip

In this paper, we  analyze gravitational Drell-Yan production at the LHC, extending the study of \cite{lp} towards Planckian center-of-mass energies.
 The UV finiteness of the amplitude in principle offers access to the cross over scale $\Lambda_T$. We implement explicit RG equations for the running of Newton's coupling and also study the RG scheme dependence of amplitudes. We show how to relate the RG scale with the Kaluza-Klein mass and kinematic variables of the scattering process. We evaluate the amplitude both numerically and analytically, using analytical continuation in energy or principle value integration respectively.  We also explain how the form-factor approximation  and effective theory fit within our picture in specific limits.  Interestingly, we find that the Kaluza-Klein gravitons act as messengers of the higher dimensional fixed point which tend to suppress the amplitude on the brane.
 While the energy-dependence of parton-level amplitudes is folded with strongly decaying parton distribution
 functions, we still find  significant sensitivity to the scale $\Lambda_T$ over SM backgrounds.
\medskip

We organize this paper as follows: we start with a brief review of the single graviton amplitude and its significant UV sensitivity within effective theory (Sect~\ref{sec:virt}). The basics of the asymptotic safety scenario for quantum gravity are discussed (Sect~\ref{sec:qgrav}), including explicit RG equations 
for the running of Newton's coupling. We then implement quantum gravity corrections of the amplitude, replacing the perturbative graviton by its  RG improved counterpart  (Sect~\ref{sec:high}). We evaluate various schemes to relate the RG scale with the kinematics of the scattering process and the Kaluza-Klein masses. We show that the KK gravitons lead to a suppression of amplitudes at Planckian energies, which are computed using either analytical continuation or principal value integration. We also discuss the approximation of isolated graviton widths and graviton decay.
We display our  results for gravitational di-lepton production at the LHC (Sect~\ref{sec:lhc}), discuss the validity and RG scheme (in-)dependence of our approximations, and compare with effective theory. We summarize our findings (Sect~\ref{discussion}) and conclude (Sect~\ref{conclusions}).

\section{Effective theory} \label{sec:virt}
In this section we recall the basics of gravitational scattering in large extra dimensions and introduce or notation. We work in $d=4+n$ dimensions, where $n$ denotes the number of extra spatial dimensions. Furthermore, $M_*$ stands for the fundamental Planck scale of the order of a few TeV, while $M_{\rm Pl}$ denotes the four-dimensional Planck scale. 
In an $s$-channel graviton exchange
\begin{equation}
q \bar{q}, gg 
\; \stackrel{G}{\longrightarrow} \; 
\mu^+ \mu^- \; ,
\end{equation}
for given final-state kinematics any graviton in the Kaluza-Klein (KK) tower can
appear as an intermediate state. These states must be summed at the
amplitude level.  Since this KK sum involves a very large number of 
states ($\sim 10^{32}$) with mostly regular spacing well below any
experimental resolution, we replace it with an integral
\begin{equation}
\frac{1}{\mpl^2}
\displaystyle\sum_{n_1=-\infty}^{\infty}\cdots 
\displaystyle\sum_{n_n =-\infty}^{\infty} \cdots 
\quad \longrightarrow \quad 
\frac{\sd}{\mstar^{n+2}} \int d\mkk\: \mkk^{n-1}\; \cdots \; .
\label{eq:sumint} 
\end{equation} 
The surface area of an $n$-dimensional sphere gives $\sd \equiv
2\pi^{n/2}/\Gamma(n/2)$. Absorbing the geometric size of the
extra dimensions $R$ into the coupling, $\mpl = \mstar (2 \pi R
\mstar)^{n/2}$, the tower of KK states now couples proportional to
$\mstar$. 
It is instructive to go back to the full $(4+n)$-dimensional
theory, where we only have the fundamental Planck scale $\mstar$ and
the gravitons propagating in the the full (brane $+$ bulk) space.
In this case the sum leading to (\ref{eq:sumint}) is merely a loop
integral over the unfixed bulk momentum, with the appropriate powers 
of $\mstar$ in front.  The power of $\mstar$ is  fixed by the 
higher-dimensional Newton coupling, $[G_D]= -(2 +n)$. \medskip

From the KK graviton Feynman rules~\cite{grw,coll_tao} it follows 
 that for a massless Standard Model ($T_{\mu}^{\mu}=0$) and 
energy-momentum conservation ($k_{\mu} T^{\mu\nu}=0$), the tensor 
structure of the graviton matter coupling simplifies significantly.  
Tree--level graviton exchange is described by the amplitude 
${\cal A} = {\cal S}\cdot {\cal T}$, where ${\cal T} = T_{\mu\nu}
T^{\mu\nu} - \frac{1}{2+n} T_\mu^\mu T_\nu^\nu$ is a function of the
energy-momentum tensor, and
\begin{equation}
\s(s)\equiv\frac{\sd}{M_*^{n+2}} \int dm \: 
\frac{m^{n-1}}{s-m^2+i\epsilon} \; .
\label{amplitude}
\end{equation}
defines the summed KK kernel, using~(\ref{eq:sumint}). The amplitude \eq{amplitude} is ultraviolet divergent for $n \ge 2$, infrared divergent for $n <0$ and logarithmically divergent in $n =2$. For $0<n <2$ the integral \eq{amplitude} converges.  By dimensional analysis the amplitude would behave as
\begin{equation}
\label{S-DR}
{\cal S}(s)=-\frac{S_{n -1}}{M_*^4}\left(\frac{-s}{M_*^2}\right)^{n /2-1}\,C\,,
\end{equation}
but the numerical coefficient $C$ is divergent. A physical meaning can be given to the amplitudes for $n >2$ by analytical continuation of the result for $0<n <2$, corresponding to dimensional regularization. In this case one finds $C_{\text{DR}}=\pi^{n /2}\Gamma\left(1-\frac{n }{2}\right)$.  For odd dimensions $C_{\text{DR}}$ is finite, but remains divergent for even dimensions. The regularized version of the effective operator takes the form 
\begin{equation}
\label{effective}
\s=-\frac{S_{n -1}}{M_*^4}\left(\frac{\Lambda}{M_*}\right)^{n -2}C
\end{equation}
for sufficiently small $s/M^2_*\ll 1$ with
\begin{equation}
\label{Creg}
C=
\int_0^\infty dx \,x^{n-3}\, F(x)\,.
\end{equation}
Here $F(x)$ denotes the regularization for the integral and the coefficient $C$ now parameterizes the effective operator in terms of the (unknown) UV regularization of the theory.  There are 
several ways to implement a UV cutoff~\cite{grw,coll_joanne,tevatron_virt,coll_maxim,Giudice:2003tu,thick_brane}. For a sharp cutoff the UV modes are suppressed by hand above $m=\Lambda$ and $F_{\rm sharp}(x)=\theta(1-x)$. This leads to 
\begin{equation}
\label{Csharp}
C_{\rm sharp}=\frac{1}{n -2}\,.
\end{equation}
Note that the strong UV divergence of the single graviton amplitude \eq{amplitude} is now parameterized by the strong sensitivity of \eq{effective} on the UV cutoff scale $\Lambda$.  Note additionally that the logarithmic divergence at $n =2$ becomes a simple pole in the coefficient \eq{Csharp}.\medskip

On dimensional grounds and to lowest order in $s\ll\mstar$, the effective operator representing $s$-channel graviton exchange can also be written as
\begin{equation}
\label{fit}
\s=-\frac{4\pi}{M^4_{\rm eff}}\,.
\end{equation}
This parameterization using the scale $M_{\rm eff}$ is commonly applied in the literature. In general, the mass scale $M_{\rm eff}$ cannot be determined from low-energy considerations alone, but requires input from the full quantum dynamics at the Planck scale. 
The regularization \eq{Creg} also fixes the mass scale $M_{\rm eff}$ in \eq{fit} as
\begin{equation}
\label{Meff}
\left(\frac{M_*}{M_{\rm eff}}\right)^4=\frac{S_{n -1}}{4\pi}\left(\frac{\Lambda}{M_*}\right)^{n -2}\,C\,.
\end{equation}
Below, we will denote the effective theory cutoff on the KK masses as $\Lambda_{\rm kk}$. The energy integration over \eq{amplitude} also requires a cutoff in effective theory, which we denote as $\Lambda_s$.  For the remainder of this paper we study how quantum gravity regularizes the amplitude \eq{amplitude} dynamically, provided that the high-energy behavior of gravity is governed by a renormalization group fixed point.

\section{Quantum gravity}
\label{sec:qgrav}
In this section, we discuss the asymptotic safety scenario for gravity and adopt it for the purposes of the present paper.  It is well known that the standard perturbative quantization of gravity faces problems.  In this section we discuss S.~Weinberg's scenario known as asymptotic safety \cite{weinberg} (see \cite{Litim:2006dx,Niedermaier:2006ns,Percacci:2007sz,Litim:2008tt}
 for reviews), where the high energy behavior of gravity is non-perturbative and dictated by an interacting fixed point under the renormalization group.   We also introduce some of our notation and discuss the main implications as well as explicit renormalization group equations for gravity. 

\subsection{Asymptotically safe gravity}
As in any non-trivial quantum field theory, once quantum fluctuations of the propagating field -- the metric field --  are taken into account, the corresponding (Newton's) coupling $G_N$ becomes a `running' coupling as a function of the renormalization group scale $\mu$. To set the stage, we will assume a general $d$-dimensional gravitational action including matter couplings.  We are interested in the coupling of Standard Model particles living on a four-dimensional brane with $d$-dimensional gravity.  The fundamental Planck scale $M_*$ is of the order a few TeV. Within effective theory, quantum gravitational corrections can reliably be computed for characteristic energies below the fundamental Planck scale $E^2\ll M^2_*$, provided that the theory is equipped with a UV cutoff. We write the renormalized  (running) gravitational coupling as
\begin{equation}
\label{G}
 G(\mu) = G_N\, Z^{-1}(\mu)\,,
 \end{equation}
where $\mu$ denotes the RG scale. We have also introduced the wave function renormalization factor $Z(\mu)$ accounting for the scale (or energy) dependence of Newton's coupling. The energy-dependence of \eq{G}  is given by the Callan-Symanzik equation for the running of Newton's coupling \cite{Litim:2006dx}
\begin{equation}
\label{dG}
\frac{d G(\mu)}{d\ln \mu}=\eta\,G(\mu)\, .
\end{equation}
Here $\eta$ denotes the graviton anomalous dimension $\eta(\mu)=-\frac{d\ln Z(\mu)}{d\ln\mu}$. Note that the anomalous dimension is a function of all couplings in the theory including the gravitational couplings, the gravity-matter couplings, and the cosmological constant. Once $\eta=\eta(\mu)$ is known explicitly as a function of the RG scale $\mu$ we have
\begin{equation}
\label{Zfrometa}
Z^{-1}(\mu)=Z^{-1}(\mu_0)\,\exp\left(\int_{\mu_0}^\mu dk \,\eta(k)\right).
\end{equation}
At low energies $\mu\ll M_*$, the quantum corrections are weak, and can be computed perturbatively provided $G \mu^{d-2}\ll 1$.  To leading order, the anomalous dimension obeys $|\eta| \ll 1$ in the perturbative regime.  However, once energies approach the Planck scale, quantum corrections become relevant and the graviton anomalous dimension becomes large $|\eta|\sim {\cal O}(1)$. In this regime the sensitivity to the high energy structure of gravity is strong, and an effective theory description is no longer applicable. \medskip

Within S.~Weinberg's scenario of asymptotically safe gravity, the domain of validity of \eq{G} is extended to energies at and above the Planck energy.
This regime is central for collider signatures of low-scale quantum gravity investigated in this paper. Within the asymptotic safety scenario it is stipulated that the metric field remains the main carrier of the gravitational force even at high energies. This is achieved provided that the relevant gravitational couplings display a non-trivial UV fixed point. The fixed point tames the virulent UV divergences of standard perturbation theory and implies a weakening of gravitational interactions at short distances, sometimes referred to as `anti-screening'. Implications of the fixed point become visible once we introduce dimensionless variables as appropriate for fixed point studies. We define the dimensionless gravitational coupling as $g(\mu) = G(\mu)\mu^{d-2}$. The corresponding Callan-Symanzik equation reads
\begin{equation}
\label{dg}
\frac{dg}{d\ln \mu}=(d-2+\eta)g(\mu)\,.
\end{equation}
From \eq{dg} we conclude that the gravitational coupling displays two types of fixed points,  a free theory fixed point at $g_*=0$ and an interacting fixed point with 
\begin{equation}
\label{etafix}
\eta_*=2-d\,.
\end{equation}
The free theory fixed point is the so-called Gaussian one. In its vicinity the anomalous dimension is small and hence the running coupling $G(\mu)\approx G_N$  is not running at all. The Gaussian fixed point governs the regime of classical general relativity. The second fixed point -- an implicit one -- requires that the anomalous dimension $\eta$ counter-balances the canonical dimension of Newton's coupling.  It is interesting to compare this behavior with standard perturbation theory in Newton's coupling. The effective expansion parameter is the dimensionless quantity $G_N E^{d-2}$, where $E$ denotes the energy scale. While this coupling is small for low energies, it grows with increasing energy, ultimately causing a breakdown of standard perturbation theory. In contrast, the dimensionless coupling at high energies remains bounded within asymptotically safe gravity and achieves a fixed point.  When expressed in terms of the renormalized Newton's coupling, the fixed point scaling leads to the behavior $G(\mu)\approx g_*/\mu^{d-2}$ for large $\mu\gg M_*$. Hence gravity is anti-screening and weakens in the onset of fixed point scaling.  \medskip

Next we discuss implications of the RG running at the level of the graviton propagator \cite{Litim:2006dx,Niedermaier:2006ns,Percacci:2007sz,Litim:2008tt}.  Neglecting tensor indices, the $d$-dimensional graviton propagator $\Delta(p^2)$ behaves as $\sim 1/p^2$ in the perturbative regime where $p^2\ll M_*^2$. RG corrections are obtained via the wave-function renormalization factor, leading to $\Delta(p^2)\to 1/(Z(\mu) p^2)$. Evaluating the RG at the scale set by the graviton momentum $\mu^2\approx p^2$ implies that the anomalous dimension modifies the propagator $\Delta(p^2) \to (p^2)^{-1+\eta/2}$. Clearly, for small anomalous dimension $|\eta|\ll 1$, the propagator remains unchanged. In turn, the propagator behaves as $\Delta(p^2)\sim (p^2)^{-d/2}$ in the fixed point regime where the anomalous dimension grows large. Hence, the renormalization group improvement of the propagator due to the dressing with virtual gravitons leads to an enhanced suppression for large external momenta  \cite{Litim:2003vp,Fischer:2006fz}. Below, we exploit the onset of fixed point scaling in the regime where center-of-mass energies approach the fundamental Planck scale.

\subsection{Functional renormalization}
Explicit RG equations for gravity -- such as \eq{dG} and \eq{dg}, and similar for other couplings --  are studied in \cite{Litim:2003vp,Fischer:2006fz,Reuter:1996cp} using the functional renormalization group \cite{Wetterich:1992yh,Litim:1998nf}. This technique is based on a Wilsonian cutoff for the propagating modes. Diffeomorphism  symmetry is controlled with the help of the background field method \cite{Reuter:1996cp,Freire:2000bq}. In order to illustrate the main result, and also for later use (see Sect.~\ref{sec:high} and ~\ref{sec:lhc}), we discuss the RG equations for the gravitational coupling in the absence of a cosmological constant and matter couplings. Following \cite{Litim:2003vp}, the RG equation for $g$ is given by \eq{dg},
where the anomalous dimension is
\begin{equation}
\label{eta}
\eta(g)=(2-d)\frac{2(d+2)g}{1-2(d-2)g}\,.
\end{equation}
We have rescaled $g$ as $g\to g/c_d$ with $c_d=(4\pi)^{d/2-1}\Gamma(\frac d2+2)$ to simplify the expressions. Note that \eq{eta} is non-perturbative in the running coupling $g$. The RG equations \eq{dg} and \eq{eta} are integrated to give the explicit solution
\begin{equation}
\label{solution}
\frac{\mu}{\mu_0}=
\left(\frac{g(\mu)}{g_0}\right)^{1/\theta_{\rm G}}
\left(
\frac{g_*-g(\mu)}{g_*-g_0}\right)^{-1/\theta_{\rm NG}}\,.
\end{equation}
The parameter $\theta_{\rm G}$ and $\theta_{\rm NG}$  are the scaling exponents at the Gaussian (G) fixed point $g_*=0$  and the non-Gaussian (NG) fixed point $g_*\neq 0$, with $g_0=g(\mu=\mu_0)$. The former is given by the canonical mass dimension of Newton's coupling in $d$ dimensions, whereas the latter is determined from the UV dynamics of the theory \cite{Litim:2003vp} with
\begin{equation}
\label{theta}
\theta_{\rm G}=d-2\,,\quad\quad \theta_{\rm NG}=2d\frac{d-2}{d+2}\,.
\end{equation} 
As a function of the RG scale $\mu$, \eq{dg} and \eq{eta} displays a cross-over from classical scaling to fixed point scaling. The scaling indices $\theta$ parameterize how rapidly each gravitational coupling and its anomalous dimension cross-over from their classical to fixed point values on a logarithmic energy scale. Note that both indices grow with dimension. In particular, the larger $\theta$, the faster the cross-over regime in units of the RG `time' $\ln \mu$. Furthermore, $\theta_{\rm NG}$ is larger than $\theta_{\rm G}$ and the ratio $\theta_{\rm NG}/\theta_{\rm G}=2d/(d+2)$ ranges between $[4/3,1/2]$ for $d$ between $[4,\infty]$. It follows that the cross-over region between IR and UV scaling narrows with increasing dimension \cite{Litim:2003vp,lp}. 

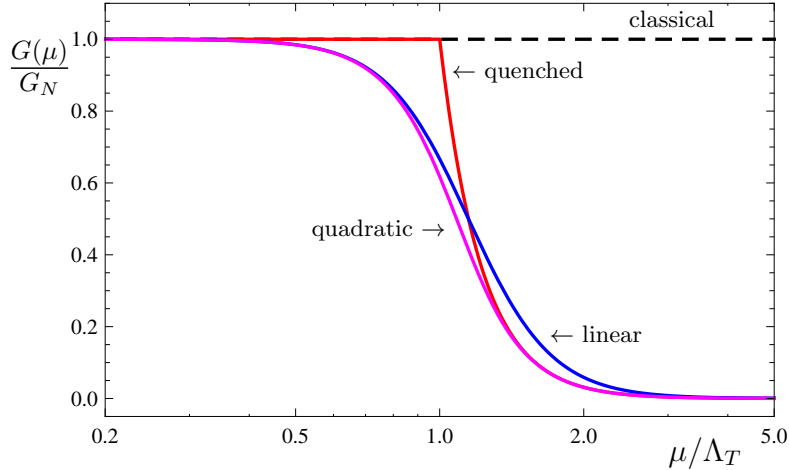
\begin{figure}[t]
\centering
\includegraphics[scale=.9]{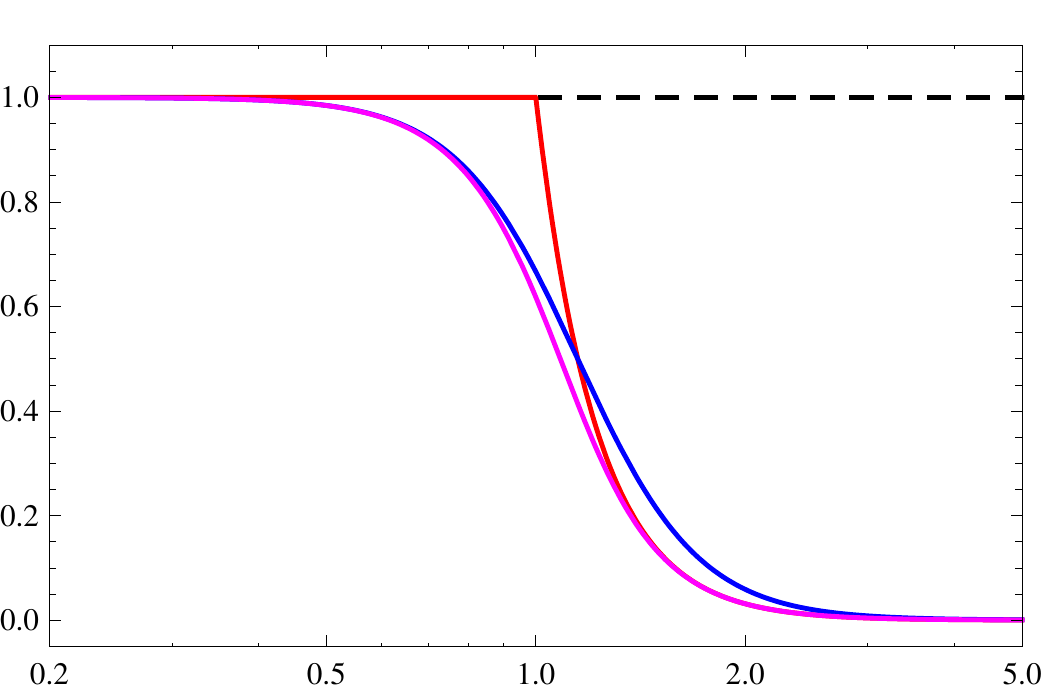} 
\put(-60,159){classical}
\put(-90,40){$\leftarrow$ linear}
\put(-127,140){$\leftarrow$ quenched}
\put(-180,80){quadratic $\rightarrow$}
\put(-295,140){\Large{$\frac{G(\mu)}{G_N}$}}
\put(-45,-5){{\large{$\mu/\Lambda_T$}}}
\caption{Crossover of the gravitational coupling $G(\mu)/G_N$ from classical (dashed line) 
to fixed point scaling (full lines) in the Einstein Hilbert theory with $n =3$ extra dimensions: 
classical behavior (black), linear crossover \eq{linear} (blue), quadratic 
crossover \eq{quadratic} (magenta) and 
the quenched approximation \eq{quench} (red) with $\Lambda_T=\Lambda^{(0)}_T=\Lambda^{(1)}_T=\Lambda^{(2)}_T$
(see text).}
\label{fig:crossoverG}
\end{figure}

\subsection{Running Newton's coupling}
\label{anomalous}
For the applications below it is useful to have explicit expressions for the functions $g(\mu), G(\mu), Z(\mu)$, and $\eta(\mu)$ at hand. We introduce three different approximations capturing the essential features.  Since the transition from perturbative to fixed point scaling becomes very narrow with increasing dimension, the scale dependence is well-approximated by an instantaneous  jump in the anomalous dimension at the energy scale $\mu=\Lambda^{(0)}_T$. This is the quenched approximation, solely characterized by a transition scale of the order of the fundamental Planck scale $M_*$, where
\begin{equation}
Z^{-1}(\mu)=
\left\{
\begin{array}{cc}
1                &{\rm for}\ \mu <\Lambda^{(0)}_T\\
(\Lambda^{(0)}_T)^{d-2}/\mu^{d-2}&{\rm for}\ \mu \ge\Lambda^{(0)}_T
\end{array}
\right.
\quad\quad
\eta(\mu)=
\left\{
\begin{array}{cc}
0&{\rm for}\ \mu <\Lambda^{(0)}_T\\
2-d&{\rm for}\ \mu \ge\Lambda^{(0)}_T
\end{array}
\right. .
\label{quench}
\end{equation}
In the quenched approximation, continuity in $G(\mu)$ follows if the fixed point value $g_*$ is related to Newton's coupling at low energies and the transition scale via $g_*=G_N (\Lambda^{(0)}_T)^{d-2}$.  We will identify the cross-over scale $\Lambda_T$ with the parameter $\Lambda^{(0)}_T$ in \eq{quench}. \medskip

In order to achieve a continuous transition we resolve \eq{solution} with mild additional approximations for the ratio $\theta_{\rm G}/\theta_{\rm NG}$. Imposing that $\theta_{\rm G}/\theta_{\rm NG}=1$, we find an explicit expression for the  running coupling \eq{G} with
\begin{equation}
\label{linear}
 Z(\mu)=1+\,G_N \mu^{d-2}/g_* \,,\quad\quad \eta(\mu)=(2-d)\left(1-\frac{1}{Z(\mu)}\right)\,.
\end{equation}
We have taken the limit $g_0\ll g_*$ and $\mu_0\ll \Lambda_T$ while keeping the gravitational coupling $G_N=g_0/\mu_0^{d-2}$ fixed to its $d$-dimensional value.  We have also used $g_*=G_N (\Lambda^{(1)}_T)^{d-2}$. Here, the anomalous dimension is linear in the dimensionless coupling $g$ and reads $\eta=(2-d)g/g_*$. Provided that   $(2-d)/g_*$ is identified with the one-loop coefficient from perturbation theory in the presence of a UV cutoff, \eq{linear} becomes equivalent to a one-loop approximation. We refer to \eq{linear} as the linear approximation.  \medskip

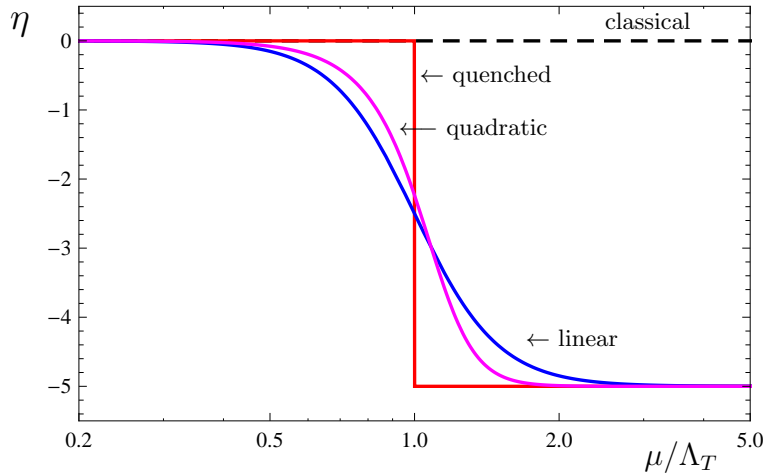
\begin{figure}[t]
\includegraphics[scale=.9]{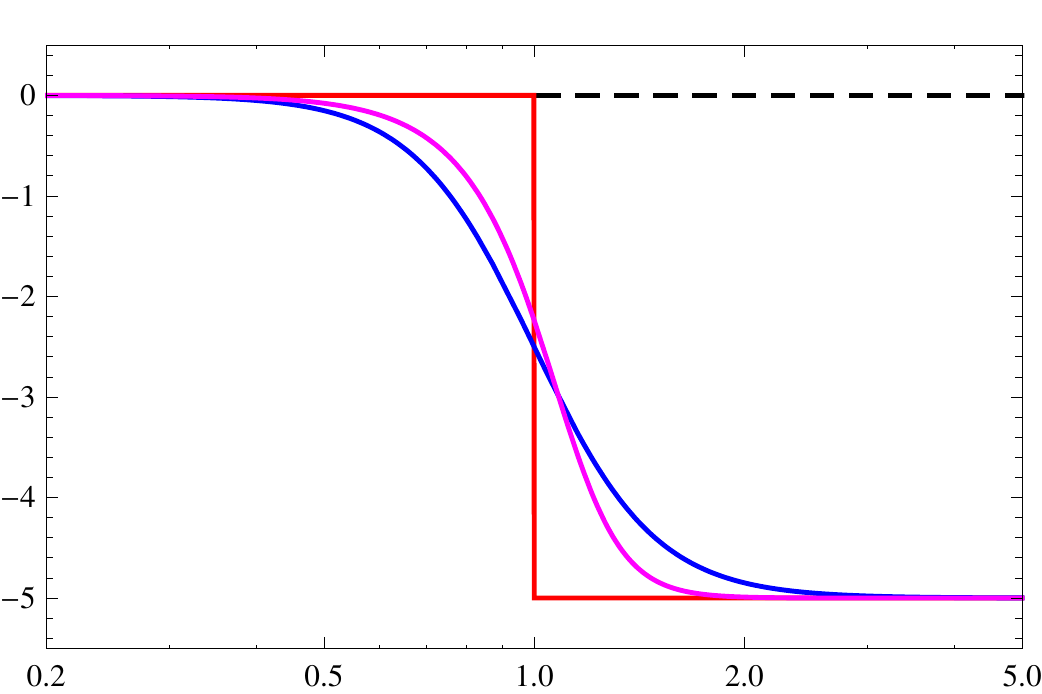} 
\put(-60,160){classical}
\put(-90,40){$\leftarrow$ linear}
\put(-130,140){$\leftarrow$ quenched}
\put(-138,120){$\longleftarrow$ quadratic}
\put(-285,160){\Large{$\eta$}}
\put(-45,-5){{\large{$\mu/\Lambda_T$}}}
\caption{Crossover of the graviton anomalous dimension from classical (dashed) to fixed point scaling (full lines) in the Einstein Hilbert theory with $n =3$ extra dimensions: classical behavior (black), linear crossover \eq{linear} (blue), quadratic crossover \eq{quadratic} (magenta) and the quenched approximation \eq{quench} (red), with $\Lambda_T=\Lambda^{(0)}_T=\Lambda^{(1)}_T=\Lambda^{(2)}_T$  (see text).}
\label{fig:crossover}
\end{figure}

Finally, a non-perturbative expression for the running coupling is achieved in the approximation $\theta_{\rm NG}/\theta_{\rm G}=2$, meaning that the approach towards the UV fixed point is twice as fast as the approach towards the Gaussian fixed point. This approximation becomes exact with increasing $d$, see \eq{theta}. The anomalous dimension and the wave function renormalization factor $Z$ are given by
\begin{equation}
\label{quadratic}
Z^{-1}(\mu)=\sqrt{1+\left(\frac{G_N}{2g_*\mu^{2-d}}\right)^2} -\frac{G_N}{2g_*\,\mu^{2-d}}\,, \quad \eta(\mu)=(2-d)\frac{Z^2(\mu)-1}{Z^2(\mu)+1}\,.
\end{equation}
Since $\eta=(2-d)g/(2g_*-g)$ in terms of the dimensionless coupling $g$, \eq{quadratic} is non-perturbative.  The fixed point value is defined as $g_*=G_N (\Lambda^{(2)}_T)^{d-2}$. We refer to \eq{quadratic} as the quadratic approximation. Note that the cross-over width in \eq{quadratic} is smaller than the one in the linear approximation \eq{linear}. More generally, for any dimension, the cross-over width of the solution \eq{solution} is bounded by the width of the linear cross-over \eq{linear} and the width of the quadratic approximation \eq{quadratic}, see Fig.~Fig.~\ref{fig:crossoverG} and \ref{fig:crossover}. 

\subsection{Relevant scales}\label{scales}
The scales relevant in the above setup are the $d$-dimensional Planck mass $M_*$, obtained from the $d$-dimensional Newton coupling via $M_*^{d-2}=1/G_N$, and the scale $\Lambda_T$ where the gravitational coupling cross over from classical to fixed point scaling. The scale $M_*$ is set by the infrared behavior of the $d$-dimensional theory, $G_N\equiv G(\mu=0)$ . In turn, the scale $\Lambda_T$ is set by the RG dynamics of the gravitational sector in the ultraviolet limit. Therefore we view  $\Lambda_T$ as  the dynamical Planck scale. We expect that $\Lambda_T$ is of the order of $M_*$, provided that the RG effects induced up to the scale $\mu\approx \Lambda_T$ are moderate. Due to the finite width of the cross-over, the scales $\Lambda^{(0)}_T$, $\Lambda^{(1)}_T$, and $\Lambda^{(2)}_T$ appearing in the approximated RG equations \eq{quench}, \eq{linear}, and \eq{quadratic} can be numerically different from the physical cross-over scale $\Lambda_T$.  The quantitative relation we work out in Sect.~\ref{sec:high}. \medskip

In scenarios where many particles couple to gravity, the externally accelerated RG running can lead to a large separation between $\Lambda_T$ and $M_*$.  We will not be concerned with these scenarios in this paper. Therefore, in the approximations \eq{quench}, \eq{linear}, and \eq{quadratic} the fixed point value $g_*$ parameterizes the uncertainty in the determination of the dynamical Planck scale $\Lambda_T$.   We also note that an analogous pattern is observed when RG improving the physics of black holes~\cite{Falls:2010he}. The RG running leads to a dynamical mass scale $M_c$ below which black holes cease to exist. Again, $M_*$ and $M_c$ are related by an order one RG factor \cite{Falls:2010he}.

\section{Virtual graviton exchange}
\label{sec:high}

In this section, we explain how quantum gravity effects are implemented via a renormalization group improvement of the single graviton amplitude.   

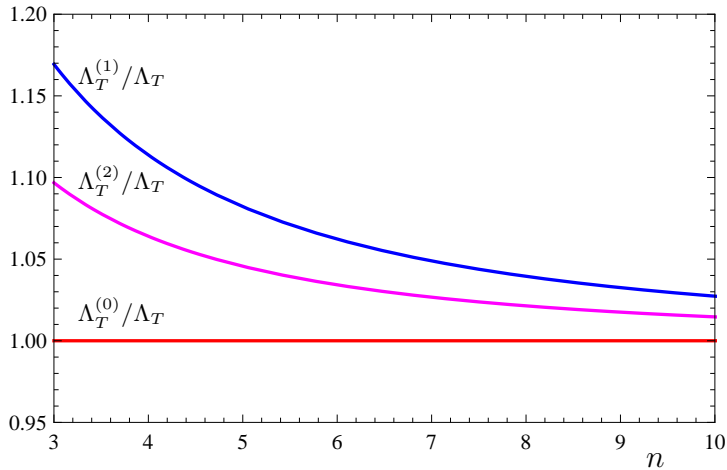
\begin{figure}[t]
\includegraphics[scale=.9]{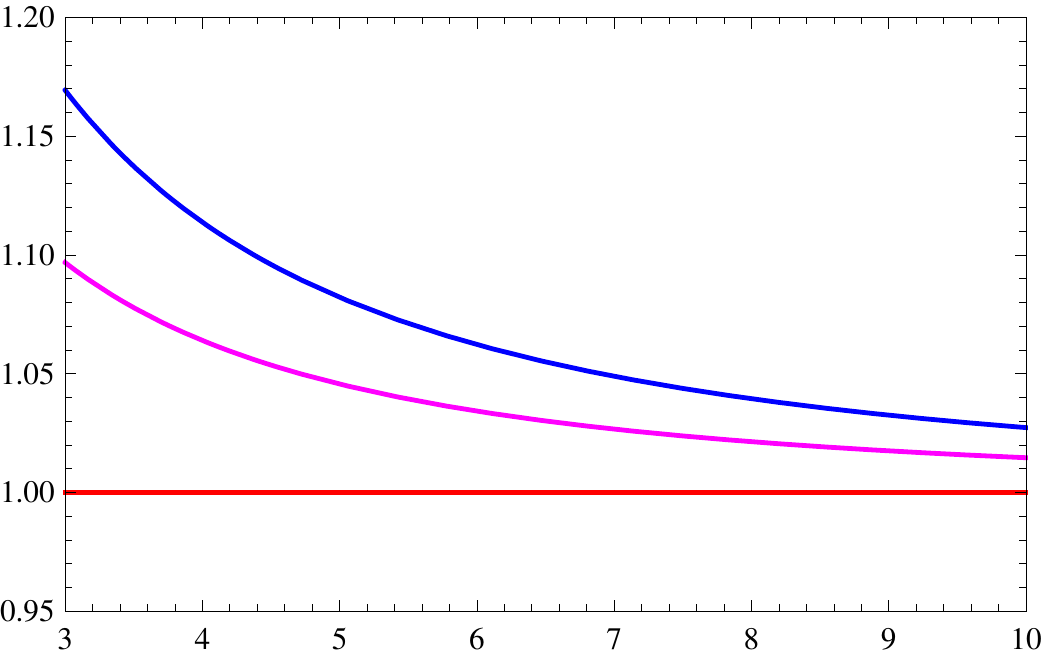} 
\put(-245,100){$\Lambda^{(2)}_T/\Lambda_T$}
\put(-245,140){$\Lambda^{(1)}_T/\Lambda_T$}
\put(-245,50){$\Lambda^{(0)}_T/\Lambda_T$}
\put(-30,-5){{\large{$n$}}}
\caption{The link between the fundamental cross-over scale $\Lambda_T$ and the RG parameters $\Lambda^{\rm RG}_T$ appearing in the quenched ($\Lambda_T^{(0)}$, red), linear ($\Lambda_T^{(1)}$, blue) and quadratic ($\Lambda_T^{(2)}$, magenta) approximations. The scale parameters differ at the 10\% level  for $n=3, 4$, and rapidly approach each other with increasing $n$.}
\label{fig:normalization}
\end{figure}

\subsection{Scale identification}
\label{scale_iden}
The tree-level single graviton amplitude in extra dimensions includes an integral over perturbative KK propagators.  Quantum gravitational corrections are included by replacing perturbative propagators with quantum corrected running propagators, and perturbative vertex functions with running vertex functions. Our key assumption is that quantum gravity corrections are primarily accounted for by replacing the perturbative propagator by the RG improved one. This step requires that the RG scale $\mu$ is identified with the scale relevant for the physical process under investigation  
\begin{equation}
\label{match}
\mu=\mu(\sqrt{s},m),
\end{equation}
in general depending on some combination of the center-of-mass energy $\sqrt{s}$ and the KK mass $m$.  A discussion of a matching which involves several scales is given in \cite{Falls:2010he} in the context of black hole production. The RG improved amplitude takes the form
\begin{equation}
\label{RGamplitude}
\s(s)=-\frac{S_{n-1}}{M_*^4}\left(\frac{\Lambda_T}{M_*}\right)^{n-2}\,C(s;\Lambda_T),
\end{equation}
where 
\begin{equation}
\label{Cgeneral}
C(s;\Lambda_T)=\int_0^\infty dx\frac{x^{n-1}}{-s/\Lambda_T^2+x^2}\,Z^{-1}\left(\mu=\mu(\sqrt{s},m,\Lambda_T)\right) ,
\end{equation}
and $x=m/\Lambda_T$. The energy dependence of \eq{Cgeneral} is encoded in the wave function factor $Z(\mu)$, itself defined via the running of the gravitational coupling \eq{G} and \eq{dG}. 
Note also that $Z(\mu)$ depends on $\Lambda_T$ through the fixed point and the $d$-dimensional Planck scale $M_*$, see Sect.~\ref{anomalous}. \medskip

Now we compare different approximations for the RG improved couplings at a given cross-over scale $\Lambda_T$. The definition of $\Lambda_T$ is fixed through the low-energy amplitude $\s(s=0)$. In particular, this implies that the cross-over scale $\Lambda_T$ coincides with the scale $\Lambda^{(0)}_T$ set in the quenched approximation.  The definition of the corresponding scale $\Lambda^{\rm RG}_T$ in terms of $\Lambda_T$ for a smooth crossover of Newton's coupling is then obtained from
\begin{equation}
\label{norm}
(\Lambda_T)^{n-2}\,C_{\rm quench}(s=0;\Lambda_T)
=(\Lambda^{\rm RG}_T)^{n-2}\,C_{\rm RG}(s=0;\Lambda^{\rm RG}_T)\,.
\end{equation}
This uniquely fixes the scale $\Lambda^{\rm RG}_T$ in terms of  $\Lambda_T$ for any explicit wave function factor $Z^{-1}_{\rm RG}$. Quantitatively, the scales $\Lambda_T$ and $\Lambda^{\rm RG}_T$ will at best differ by a few percent (see Fig.~\ref{fig:normalization}).  \medskip

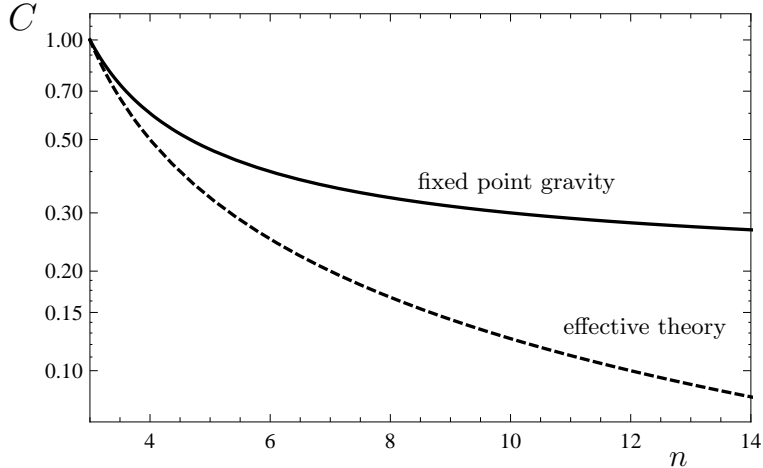
\begin{figure}[t]
\includegraphics[scale=.9]{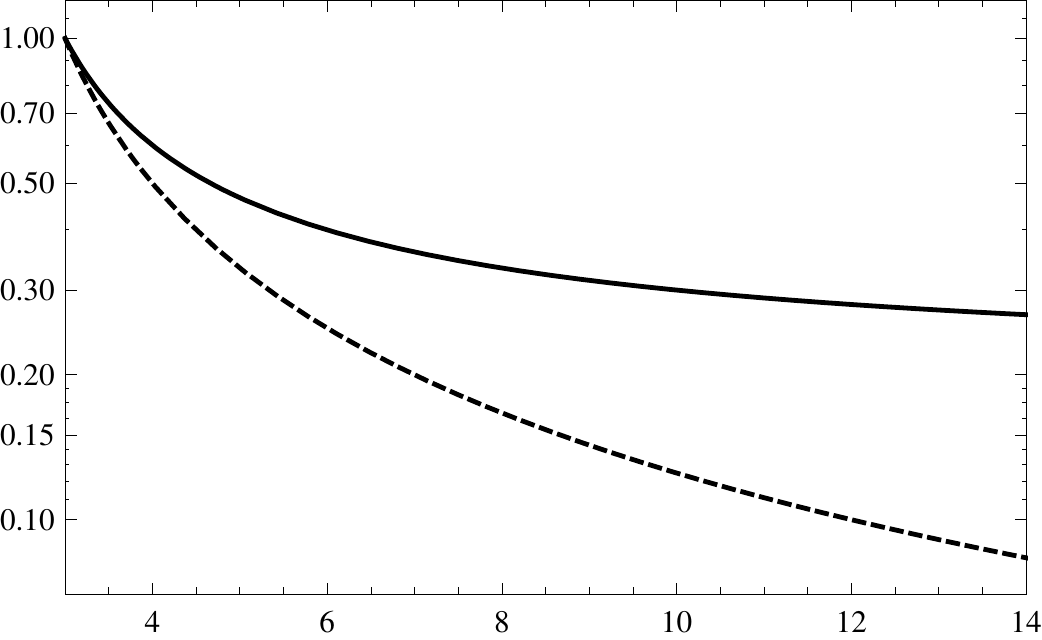} 
\put(-75,45){effective theory}
\put(-130,100){fixed point gravity}
\put(-285,160){\Large{$C$}}
\put(-35,-5){{\large{$n$}}}
\caption{Comparison of fixed point gravity with effective theory predictions for the amplitude \eq{Cgeneral} in the limit of small center-of-mass energies $s\ll \Lambda^2_T$ for various numbers of extra dimensions. The dashed line indicates the effective theory result, with a sharp cutoff on the KK mass at  $m=\Lambda_T$.
The full line represents the fixed point gravity result with a cross-over scale $\Lambda_T$. Their difference is mainly due to the UV dynamics of the KK gravitons, and becomes more pronounced with increasing $n$.}
\label{fig:coefficient}
\end{figure}

\subsection{Kaluza-Klein gravitons}\label{KKsector}
First, we analyze the effective operator for single graviton exchange \eq{RGamplitude} and \eq{Cgeneral} in the limit of small center-of-mass energy.  This approximation is motivated at the LHC by the sharply falling parton densities.  Within effective theory, the integration over the KK gravitons is cut off and we obtain the well-known result \eq{Csharp}.  Next, we introduce the  renormalization group corrections predicted by asymptotically safe gravity.  In the limit of small $\sqrt{s}\ll M_*$, the matching \eq{match} becomes
\begin{equation}
\label{matchm}
\mu=m\,,
\end{equation}
so that the RG scale is matched to the KK mass. For the coefficient \eq{Cgeneral}, we find
\begin{equation}
\label{CRG}
C=
\int_0^\infty dx \,x^{n-3}\, Z^{-1}(x\,\Lambda_T)\,.
\end{equation}
\begin{figure}[t]
\includegraphics[scale=.9]{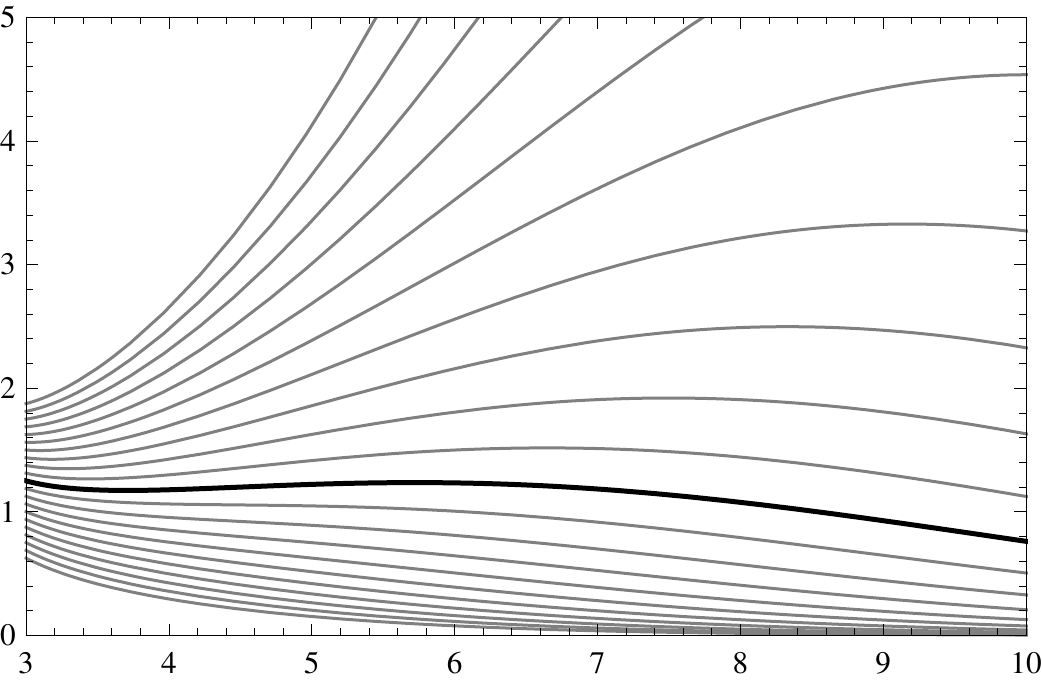} 
\put(-300,150){\large{$|\s| \,\frac{M_*^4}{4\pi}$}}
\put(-40,-15){{\large{$n$}}}
\caption{The $n$-dependence of the amplitude $|\s|\, {M_*^4}/{(4\pi)}$ defined in \eq{RGamplitude}. Thin grey lines cover $\Lambda_T/M_*=0.5$ to $1.5$ in steps of $0.05$ (from bottom to top). Close to $\Lambda_T/M_*=1$ (thick black line), the overall $n$-dependence of the amplitude is very weak.}
\label{fig:NDependence}
\end{figure}
Here, $x=m/\Lambda_T$ measures the KK mass in units of $\Lambda_T$. Note the similarity between \eq{CRG} and the regularized expression \eq{Creg} from effective theory.  Within the renormalization group, the scale-dependence of the wave function factor $Z^{-1}$ takes over the role of the regulator function $F(x)$ in \eq{Creg}. The coefficient \eq{CRG} is computed for a given renormalization group trajectory $G(\mu)=G_N Z^{-1}(\mu)$. Analytical results for \eq{CRG} are obtained in the approximations introduced in Sect.~\ref{anomalous}. In the quenched approximation \eq{quench}, we find \cite{lp}
\begin{equation}
\label{C0}
C_{(0)}=
\frac{1}{n-2}+\frac{1}{4} \equiv C_{\text{EFT}}+C_{\text{UV}}\,.
\end{equation}
We  emphasize that the first term $C_{\text{EFT}}$ is identical to the contribution within effective theory \eq{Csharp}, provided the effective theory cutoff  coincides with the energy scale where gravity becomes non-perturbative. The second term $C_{\text{UV}}$ originates solely from the KK gravitons with mass $m>\Lambda_T$ $(x>1)$. For low numbers of extra dimensions these terms are similar in magnitude. However, with increasing $n$ the contributions from the trans-Planckian domain dominate over the sub-Planckian regime. In the linear approximation \eq{linear}, we have
\begin{equation}
\label{C1}
C_{(1)}=\frac{1}{4}\Gamma \left(\frac{n-2}{n+2} \right)\,\Gamma \left(\frac{n+6}{n+2} \right)\,
= \, \frac{1}{4}+\frac{2}{3}(\frac{\pi}{n})^2 + \mathcal{O}\left(\frac{1}{n^3}\right).
\end{equation}
We see that \eq{C1} reaches the same limit as \eq{C0}. For the quadratic approximation \eq{quadratic}, we obtain
\begin{equation}
\label{C2}
C_{(2)}=-\frac{\Gamma(1+\sm{2}{n+2})\Gamma(-\sm{4}{n+2})}{(n+2)\,\Gamma(2-\sm{2}{n+2})}
\,=\,
\frac{1}{4}+\frac{1}{2n}+ \mathcal{O}\left(\frac{1}{n^2}\right).
\end{equation}
The large-$n$ limit implies that $C_{(0)}\ge C_{(2)}\ge C_{(1)}$, which holds for all $n$. Using $\Lambda_T$ within the quenched approximation as the reference point, we employ the results \eq{C0}, \eq{C1}, and \eq{C2} to identify the relative normalization of scales using \eq{norm}.  We obtain $\Lambda_T/\Lambda^{(i)}_T=(C_{(i)}/C_{(0)})^{1/(n-2)}$. This normalization implies a small shift of the order of a few percent or less, see Fig~\ref{fig:normalization}. \medskip

Next we discuss the full $n$-dependence of the amplitude \eq{RGamplitude} depending explicitly on $n$ via $S_{n-1}$, and implicitly through the RG factor $C$ and scale $\Lambda_T$ (see Fig.~\ref{fig:NDependence}). The phase space factor $S_{n-1}$ grows until $n\approx 8$, and decays for larger $n$. The coefficients \eq{C0}, \eq{C1}, and \eq{C2} decrease until $n\approx 10$ and stay at their asymptotic value thereafter. Interestingly, their product is $n$-independent for $n=3,\cdots 8$ so that the main $n$-dependence of \eq{RGamplitude} originates from the prefactor $(\Lambda_T/M_*)^{n-2}$.  Therefore, the main dependence is on the value of $\Lambda_T$ in $d=4+n$ dimensions. \medskip

In Fig.~\ref{fig:coefficient}, we compare the production amplitude \eq{Cgeneral} at small $s\ll \Lambda_T^2$, parameterized by coefficient $C$ from effective theory and the renormalization group.  Within effective theory, the IR dynamics of KK gravitons are retained but their UV dynamics are suppressed due to a cutoff at $m=\Lambda_T$.  The amplitude becomes very small for large $n$.  \medskip

Within asymptotically safe gravity, 
the UV dynamics of KK modes is controlled by a fixed point for the gravitational couplings. Quantitatively, the UV contributions to the amplitude dominate with increasing $n$. The amplitude remains bounded and approaches a finite universal limit for large $n$ as  a consequence of the underlying fixed point. This can be understood from \eq{CRG} as follows. For small $x<1$, the factor $Z(x\Lambda_T)\approx 1$ is bounded.  Because of the volume factor $\sim x^{n-3}$ in the KK integration, the sub-Planckian modes with $x<1$ will no longer contribute to \eq{CRG} in the limit of many extra dimensions.  In the trans-Planckian regime $x>1$ on the other hand, fixed point scaling implies $\mu\partial_\mu Z=(n+2)Z$, see \eq{G}, \eq{dG}, and \eq{eta}. Then, under the integral we can trade the RG scaling $\mu\partial_\mu $ with $x\partial_x$-scaling.  In the limit $1/n\to 0$ the scaling regime sets in at $x\ge1$ and the integrand becomes a total derivative $C=\frac{1}{4}\int^1_\infty dx\, \partial_x(x^{-4})$, whence $C=\frac{1}{4}$ independently of the details of the RG trajectory $Z(x\Lambda_T)$. The same conclusion is reached by recalling that the volume of an $n$-dimensional sphere is concentrated in the vicinity of its boundary for large $n$.  We also note that the coefficient approaches the limit  $C=\frac{1}{d-n}$ in a set-up where particles are confined to a $(d-n)$-dimensional brane. \medskip

\begin{figure}[t]
\includegraphics[scale=.9]{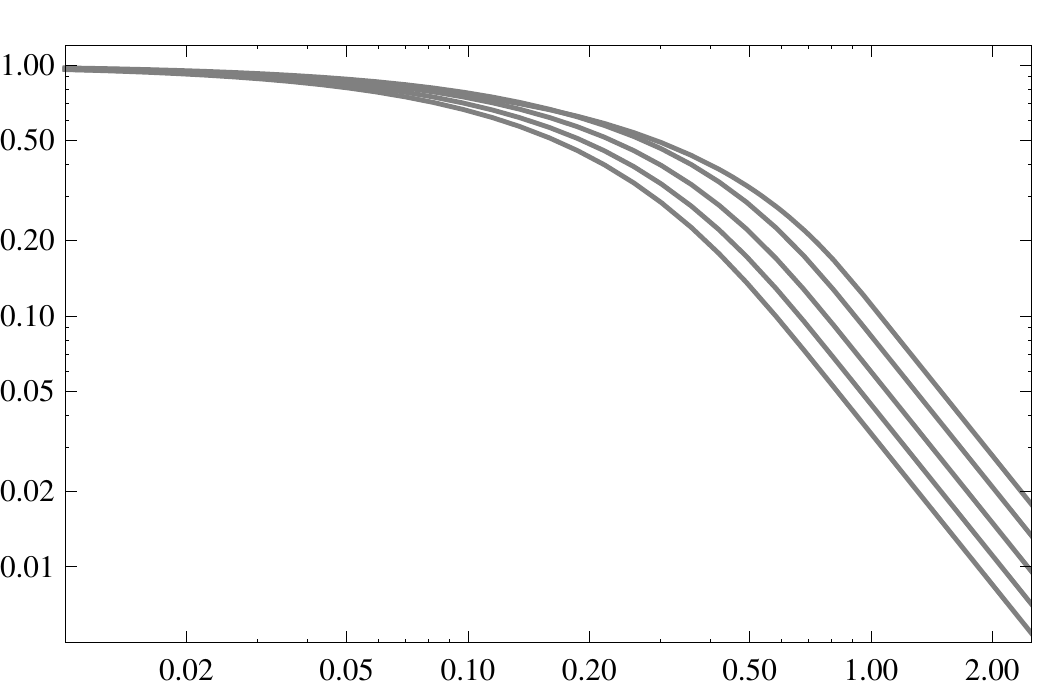} 
\put(-75,40){$n=12\ \rightarrow$}
\put(-70,130){$\leftarrow$ $n=4$}
\put(-300,150){\large{$\frac{C(s)}{C(0)}$}}
\put(-40,-15){{\large{$s/\Lambda_T^2$}}}
\caption{The energy-dependence of the functions $C(s)$ from \eq{Cs} within fixed point gravity in the quenched approximation, and $n=4,6,8,10$ and $12$ (top to bottom).  The amplitude becomes suppressed at scales below $\Lambda_T$.}
\label{fig:coefficientS}
\end{figure}
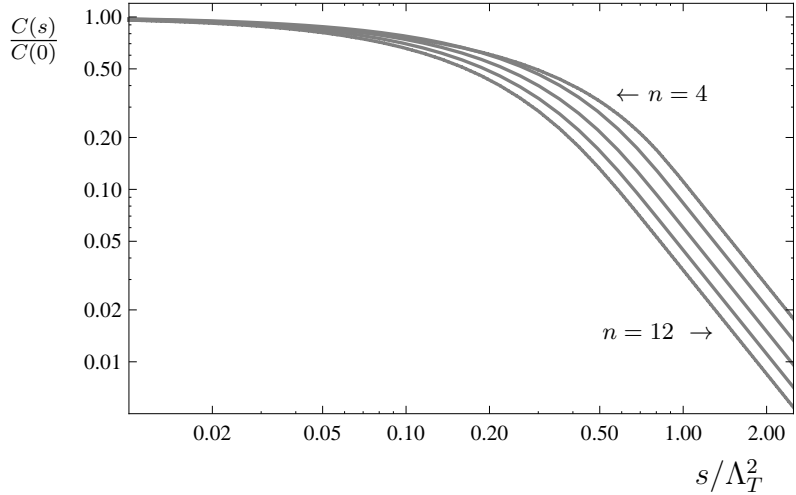

\subsection{Planckian energies}\label{Planck}
\begin{figure}[t]
\includegraphics[scale=.9]{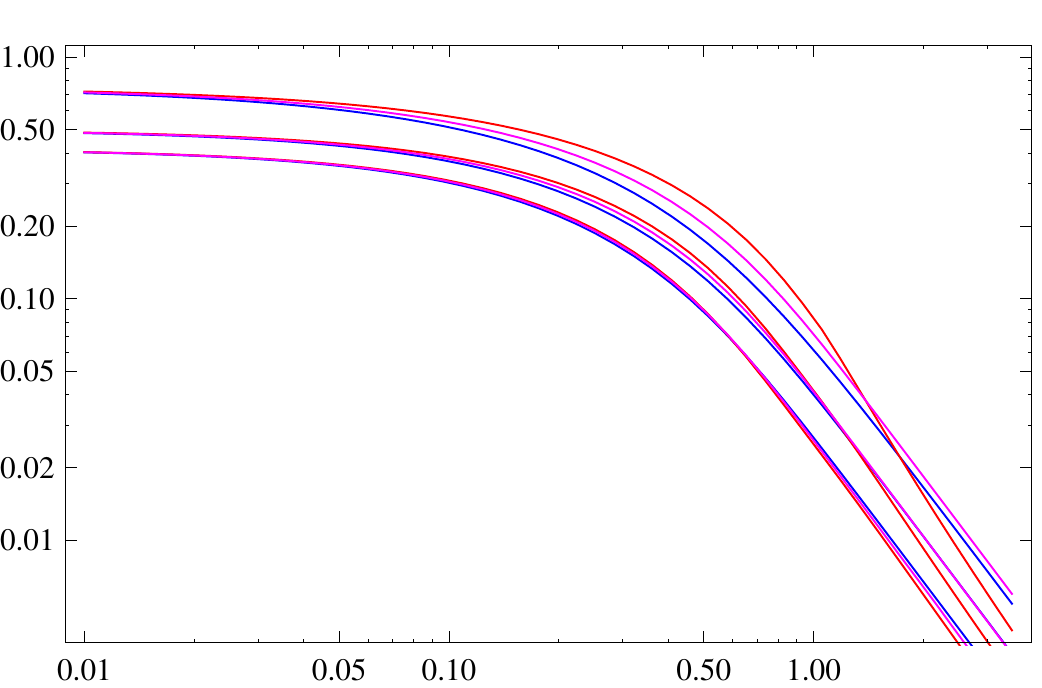} 
\put(-85,40){$n=8\ \rightarrow$}
\put(-95,60){$n=6\ \longrightarrow$}
\put(-140,150){$\leftarrow\ n=4$}
\put(-300,150){\large{$C(s)$}}
\put(-40,-15){{\large{$s/\Lambda_T^2$}}}
\caption{Scheme independence of the amplitude $C(s)$ from asymptotically safe gravity and cross-over scale $\Lambda_T$ for  $n=4, 6$ and $8$ (top to bottom), comparing the quenched (red), linear (blue), and quadratic (magenta) approximations. With increasing $n$, differences between the RG schemes are washed out by the dynamics of KK gravitons. In all cases, a suppression of the amplitude sets in at scales $s$ below the scale $\Lambda^2_T$.}
\label{fig:Compare}
\end{figure}
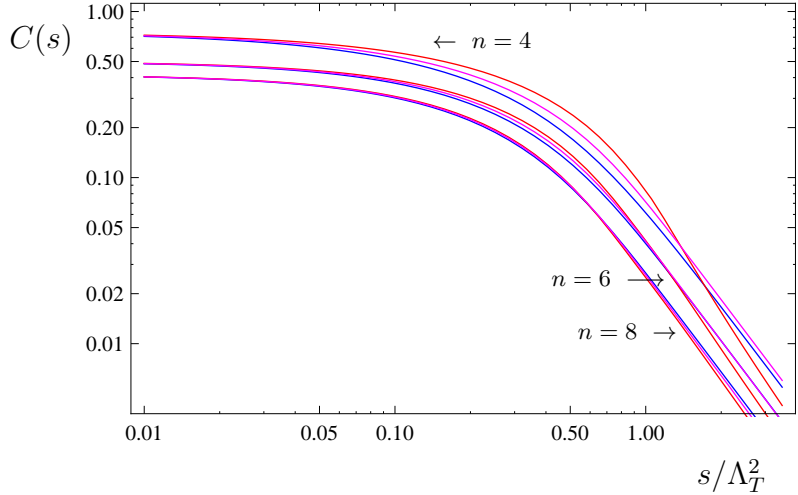

Next, we are interested in the behavior of amplitudes at Planckian center-of-mass energies. Expressions for the RG improved single graviton amplitude are obtained by identifying the renormalization group scale $\mu$ with the $d$-dimensional euclidean momentum scale,
\begin{equation}
\label{matchs+m}
\mu^2=s+m^2\,. 
\end{equation}
This identification is most natural from a $d$-dimensional perspective as the effective single graviton amplitude is sensitive to the $d$-dimensional momenta running through the propagator. We evaluate \eq{matchs+m} together with an analytical continuation to euclidean signature, and obtain  finite results for the single graviton amplitude. Using the definition for $Z(\mu)$ via \eq{G} and \eq{dG}, we find
\begin{equation}
\label{Cs}
C(s)=\int_0^\infty dx\frac{x^{n-1}}{y^2+x^2}\,Z^{-1}\left(\sqrt{y^2+x^2}\,\Lambda_T\right)\,\quad{\rm with}\quad y^2=s/\Lambda_T^2\,.
\end{equation}
We emphasize that the $s$-dependence of \eq{Cs} originates from both the single graviton exchange and the RG dressing of the propagator. In Fig.~\ref{fig:coefficientS}, we compute \eq{Cs} in the quenched  approximation  for various extra dimensions. For small $s/\Lambda^2_T\ll 1$ the variation compared to $C(s=0)$ is small, though slightly stronger for increasing $n$. With increasing number of extra dimensions,  the crossover from perturbative scaling to high-energy scaling sets in at lower energy scales $s<\Lambda^2_T$.  This new pattern is largely insensitive to the details of the RG running as can be seen from Fig.\ref{fig:Compare}, where we establish the RG scheme (in-)dependence by comparing the quenched, linear, and quadratic approximation. Moreover, while a small difference at $s\approx\Lambda_T^2$ remains visible for $n=4$, the results with increasing $n$ in all three approximations are barely distinguishable at all values of $s$. The pre-Planckian suppression also indicates that the RG-improved amplitudes are compatible with perturbative unitarity.
This pre-Planckian suppression in the effective operator is also noteworthy because the underlying $d=(4+n)$-dimensional running coupling,  with increasing $n$, displays an increasingly sharp crossover at the scale $\Lambda_T$. Hence, the extra-dimensional integration of KK modes implies that the transition region between classical and fixed point scaling becomes wider, and shifted towards scales below the fundamental transition scale $\Lambda_T$. This result also implies that the domain of applicability for standard effective theory becomes smaller with increasing $n$. \medskip

\subsection{Universality}
For asymptotically large $s/\Lambda_T^2\gg 1$, the amplitude \eq{Cs} is solved analytically and independent of the renormalization group trajectory $Z^{-1}(\mu)$. Since $Z$ is deep in the scaling regime for large $\sqrt{s}$ we find that $\lim_{\mu\to\infty} \mu^{n+2}Z^{-1}(\mu)$ becomes a $\mu$-independent constant, also  using \eq{Zfrometa} and \eq{eta}. Therefore, the leading behavior reads
\begin{equation}
\label{Cdecay}
C(s)= c_n\, \left(\frac{\Lambda_T^2}{s}\right)^2+{\rm subleading}\, ,
\end{equation}
with the coefficient
\begin{equation}
\label{cn}
c_n=\lim_{s\to \infty}\int_{\Lambda_T/\sqrt{s}}^\infty dz\frac{z^{n-1}}{(1+z^2)^{n/2+2}}=\frac{2}{n(n+2)}\,,
\end{equation}
where we have substituted $z=m/\sqrt{s}$. The leading large-$s$ decay \eq{Cdecay} is universal, {\it independent} of the number of extra dimensions, and fully dominated by the gravitational fixed point. All reference to the renormalization group trajectory have dropped out.  \medskip

This result can also be understood on dimensional grounds. The tree-level amplitude $\s(s)$ in \eq{RGamplitude} has canonical mass dimension $[\s(s)]=-4$. From the four-dimensional (brane) perspective it can be read as $\propto G_{\rm eff}(s)/s$, where $G_{\rm eff}(s)$ is the effective energy-dependent Newton's coupling in four dimensions. In the scaling regime $\sqrt{s}/\Lambda_T\gg 1$, $\sqrt{s}$ is the only remaining energy scale. Therefore $\s(s)\propto1/s^2$, and the number of extra dimensions is not differentiated by the scaling. 
This behavior equally implies $G_{\rm eff}(s)\propto 1/s$, which is the expected pattern for an RG fixed point in four dimensions. Hence, after the KK modes are integrated out, the amplitude $\s(s)$ behaves as if an effective four-dimensional graviton is exchanged whose high-energy behavior is governed by a four-dimensional fixed point with a cross-over scale $\Lambda_T$ parametrically smaller than the four-dimensional Planck scale $M_{\rm Pl}$. In this case, the effective four-dimensional fixed point is inherited from  the underlying higher-dimensional fixed point.  \medskip

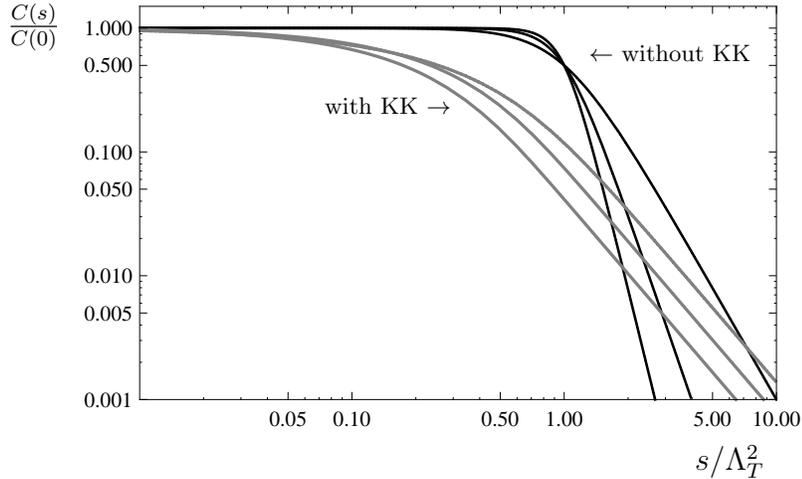
\begin{figure}[t]
\includegraphics[scale=.9]{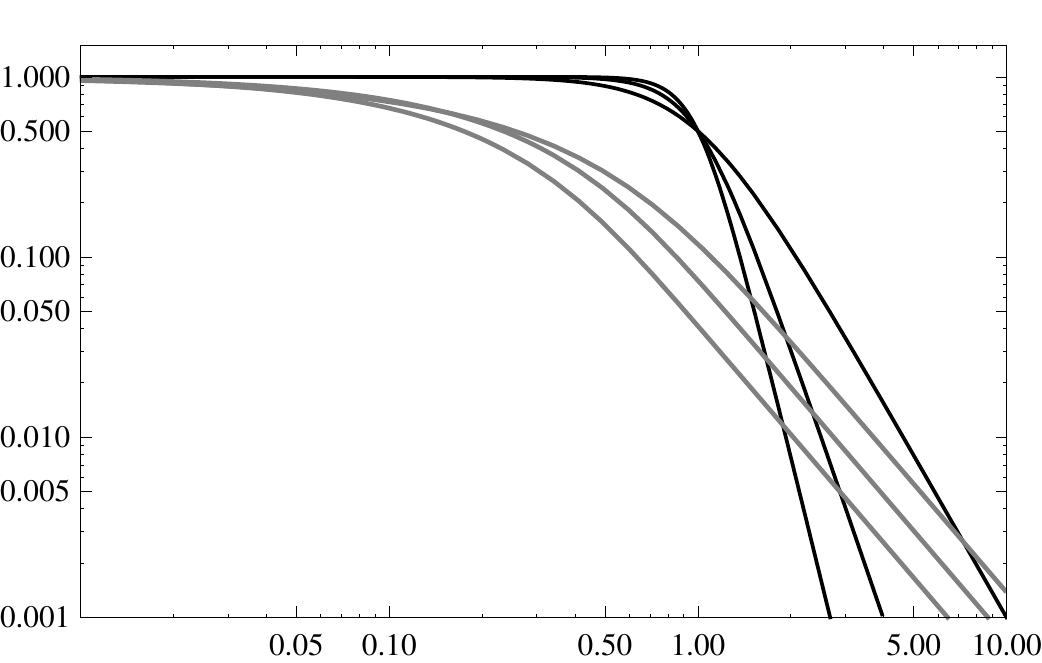} 
\put(-80,140){$\leftarrow$ without KK}
\put(-180,120){with KK $\rightarrow$}
\put(-300,150){\large{$\frac{C(s)}{C(0)}$}}
\put(-40,-15){{\large{$s/\Lambda_T^2$}}}
\caption{Impact of KK dynamics on the single graviton amplitude within asymptotically safe gravity. 
All curves are based on Newton's coupling in the linear approximation with transition scale $\Lambda_T$. 
Gray [black] lines denote the amplitudes  with \eq{Cs} [without \eq{Cfactorisation}] the UV dynamics of KK gravitons; 
$n=4,8$ and $12$  (from right to left). The inclusion of KK dynamics leads to a universal suppression of the amplitude.}
\label{fig:coefficientCompare}
\end{figure}

\subsection{Form-factor approximation}
From a four-dimensional (brane) perspective, the relevant kinematical scale is the exchanged momentum $\sqrt{s}$, suggesting the $4$-dimensional scale identification   
\begin{equation}
\label{matchs}
\mu^2=s\,.
\end{equation}
For the coefficient \eq{Cs}, this matching implies that the KK gravitons are treated perturbatively and their UV dynamics are cut off, similar to effective theory. Furthermore, the $s$-dependence originates solely from the running gravitational coupling, leading to the form-factor approximation
\begin{equation}
\label{Cfactorisation}
C(s) = Z^{-1}(\sqrt{s})\, C_{\rm KK}\,,
\end{equation}
with $Z(\mu)$ given by the renormalization group equations for gravity. Here, $C_{\rm KK}$ denotes a regularized version of the KK integral \eq{Creg}. 
\medskip

The  matching \eq{matchs} with \eq{Cfactorisation} has been adopted by  Hewett and Rizzo \cite{tom_joanne}, together with  the approximation \eq{linear} for Newton's coupling and an independent UV cutoff for the KK integration $\Lambda_{\rm kk}$. This is conceptually different from our approach (see also \cite{lp}), which does not require an additional UV cutoff. The Hewett-Rizzo parameter $t$ \cite{tom_joanne} is related to $g_*$ in \eq{linear} and the scale $\Lambda_T$ as $t=(g_*)^{1/(n+2)}=\Lambda_T/M_D$.  Note that for $t$ below $\approx 2$, it has been argued that scattering within asymptotically safe gravity is compatible with perturbative unitarity \cite{tom_joanne}. \medskip

The additional UV cutoff on the KK modes in \eq{Cfactorisation} should be set by the dynamical Planck scale. In this case, $C_{\rm KK}$ will be a similar order of magnitude as the coefficients $C$ obtained from the RG at $s=0$. In this case, we conclude that the form-factor approximation \eq{Cfactorisation} can be obtained from a matching \eq{match} provided the factorization  $Z(\mu(\sqrt{s},m))\propto Z_s(\sqrt{s})\,Z_m(m)$ holds true in the relevant kinematical regime. We also note that the $s$-dependence in \eq{Cfactorisation} originates solely from the wave function factor of the $d$-dimensional gravitational coupling. Therefore, the large-$s$ decay of the amplitude is given by the large-$\mu$ decay of the gravitational coupling, 
\begin{equation}
\label{CdecayHR}
C(s)\propto \left(\frac{\Lambda_T^2}{s}\right)^{n/2+1}\,.  
\end{equation}
With increasing dimension, the suppression becomes increasingly strong. 
In Fig.~\ref{fig:coefficientCompare}, we discuss the impact of the KK modes within asymptotically safe gravity for various $n$. The grey curves include the KK dynamics following \eq{Cs} and \eq{Cdecay}, whereas the black curves neglect their dynamics for KK masses above $\Lambda_T$,  see \eq{Cfactorisation} and \eq{CdecayHR}. Both sets of curves are based on the linear approximation \eq{linear} for the RG running of the gravitational coupling, normalized to the same strong-gravity scale $\Lambda_T$. For \eq{Cfactorisation}, the energy-dependence of the amplitude becomes visible only very close to $\Lambda_T$,  which is followed by a rapid crossover and an $n$-dependent large-$s$ decay. Furthermore, the large-$s$ behavior shows no universality if the KK dynamics are omitted. In contrast to this, fully integrating-out  the KK sector leads to a suppression of the amplitude already below $\Lambda^2_T$, and to universal scaling in the trans-Planckian regime. This summarizes the differences between our approach and the matching \eq{Cfactorisation}. We conclude that the dynamics of the KK modes is relevant already at sub-Planckian energies. The form-factor approximation  \eq{Cfactorisation}  together with \eq{matchs} is applicable for sufficiently small $s\ll M_*^2$, provided that the additional UV cutoff $\Lambda_{\rm kk}$ is set to the scale $\Lambda_T$.

\subsection{Pole region and virtuality}
\label{sec:highC}

In this section, we evaluate the KK integral in the pole region, where $s\approx m^2$ and the graviton is deemed on-shell.  
On-shell KK gravitons can also decay  into lighter particles and therefore have a non-zero decay width.  We explore the implications of the graviton width in more detail in Sec.~\ref{width} below.  Here, we will adopt the principle value prescription. In addition, we compare the matching  \eq{matchs+m} with the alternative choice 
\begin{equation}
\label{abs}
\mu^2=|s-\mkk^2|
\end{equation}
where the RG scale is identified with the virtuality of a KK graviton. Note that the main difference between the choice $\mu^2=s+m^2$ from \eq{matchs+m} and \eq{abs} relates to the pole region where $s\approx m^2$.   
Around the pole, \eq{abs} implies $\mu\approx 0$, meaning that this region is only sensitive to classical physics $G(\mu\approx 0)$, even for trans-Planckian energies $s,m^2\gg M^2_*$. In contrast, the matching \eq{matchs+m} leads to $\mu\approx m$ at the pole, which is sensitive to RG corrections.\medskip

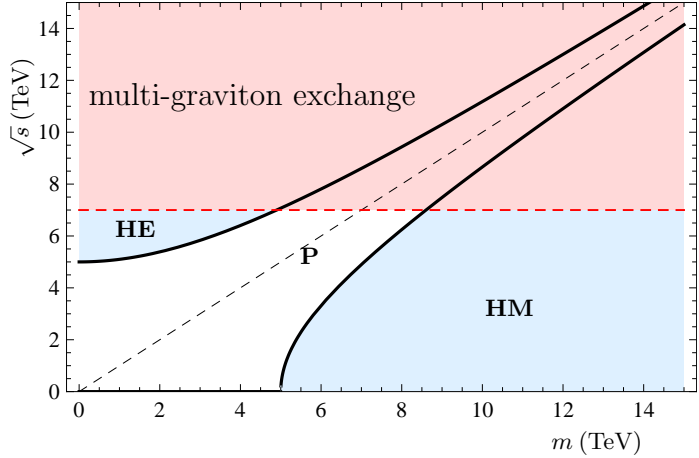
\begin{figure}[t]
\includegraphics[width=0.55\textwidth]{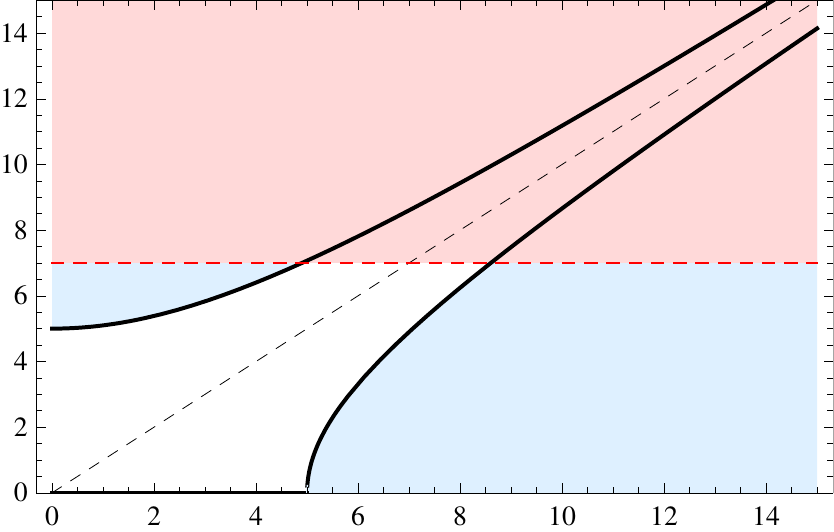} 
\put(-230,120){\large multi-graviton exchange}
\put(-220,70){\bf{HE}}
\put(-150,60){\bf{P}}
\put(-80,40){\bf{HM}}
\put(-260,100){\begin{sideways}$\sqrt{s}\:(\tev)$\end{sideways}}
\put(-55,-10){{$\mkk\:(\tev)$}}
\caption{
  UV/IR division of integration kernel using \eq{abs}
  with $\mtrans=5\,\tev$ and $\mstar=7\,\tev$.  The red region corresponds to physics
  above $s\ge \mstar^2$ (chosen for illustrative purposes), where multi-graviton and 
  width effects become relevant. KK states in the high mass (HM) and high energy (HE)
  regions are sufficiently far from their mass shell in four dimensions, 
  and thus probe the underling $4+n$ dimensional quantum gravitational
  theory. The boundaries between pole, HM and HE regions are indicative, 
  and smeared-out due to the RG running.}
\label{fig:sfunctfig}
\end{figure}

For illustration, the physics content of the matching \eq{abs} is depicted in Fig \ref{fig:sfunctfig}  for a specific choice of $\mstar$ and $\Lambda_T$.  
Single graviton exchange is reliable for $\sqrt{s}$ below the fundamental Planck scale.  For KK modes starting with masses of the order $\sqrt{s+\Lambda_T^2}$, RG corrections must be taken into account. This is denoted as the high mass region (HM) in Fig \ref{fig:sfunctfig}.  A high energy region (HE) requiring RG effects may also exist, provided that the cross-over scale $\Lambda_T$ is below the fundamental scale of gravity $\mstar$.  As the collider energy approaches $\mstar$, we expect multi-graviton processes or the production of mini-black holes to become important. Finally, we define the pole region (P), where $s\approx m^2$ and the graviton is on-shell. \medskip

We quantify the pole contribution using the RG running in the linear approximation \eq{linear}, together with \eq{abs}. The KK kernel falls off sufficiently quickly as a function of $\mkk$ and we can adopt the prescription 
\begin{alignat}{5}
Z^{-1}
\frac{m^{n-1}}{s-\mkk^2+i\epsilon}=
Z^{-1}
\,\text{P}\frac{m^{n-1}}{s-\mkk^2}+\frac{i\pi}{\sqrt{s}} \mkk^{n-1}
\delta(\sqrt{s}-\mkk)\,.
\label{ieps}
\end{alignat} 
We also use the fact that $Z^{-1}=1$ at $\mu=0$ for all approximations.
The second term in \eq{ieps} is responsible for the on-shell production of KK gravitons.  
The principal value integration we can perform analytically for all $n=2+4i$ and integer $i$. 
For $n=2$ we obtain
\begin{eqnarray}
C(s)&=&{\rm P}\int dx\frac{x}{-s+x^2}\,Z^{-1}(\mu^2=|s-x^2|\Lambda^2_T)
=\frac{1}{4}\ln\left(1+\frac{1}{s^2}\right)
\label{linear2}
\end{eqnarray}
with $x=m/\Lambda_T$, and $s\to s/\Lambda_T^2$.
We note that the amplitude is well-defined for all $s$, and the result agrees with the one obtained by analytical continuation. The amplitude decays asymptotically as $\propto 1/s^2$.  For comparison, effective theory gives $C_{\rm eff}(s)={\small \frac 12}\ln(1/s-1)$ valid for $s/ \Lambda_T^2<1$ \cite{coll_tao}, which has the same logarithmic singularity as (\ref{linear2})  for ultra-soft $s/ \Lambda_T^2\ll1$. 
Next we consider  $n=6$,
where
\begin{eqnarray}
C(s)
&=&\frac{\pi\,s}{2\sqrt{2}}+\frac{s^2}{8}\ln\left(1+\frac{1}{s^4}\right)-\left(\frac{s}{4\sqrt{2}}\ln\left(A(s)+s^2\right)
-\frac{A(s)}{4}\arctan A(s) +(s\leftrightarrow -s)\right)\,,
\label{linear6}
\end{eqnarray}
and  $A(s)=1-\sqrt{2}s$, $x=m/\Lambda_T$, while $s$ is again expressed in units of $\Lambda_T^2$.
The result agrees with \eq{Cs} after analytical continuation $s\to -s$. For large negative $s$, we 
find that the amplitude \eq{linear6} decays as $s^{-2}$. Again, this matches the coefficient
from \eq{Cdecay} in this limit. For large positive $s$, \eq{linear6} grows linearly with $s$, approaching $\pi\,s/\sqrt{2}+{\cal O}(1/s^2)$. We stress that the linearly growing term in $s$ originates from the density of states at the pole region for $s,m^2> \Lambda_T^2$. As discussed above, this behavior is an effect of the virtuality matching \eq{abs}, which treats the modes at the pole with $s,m^2\gg \Lambda_T^2$ as classical. In fact, the growing term is absent for the euclidean matching \eq{matchs+m}, which induces a suppression of the density of states for large KK masses. In comparison with effective theory, we note that the amplitude from fixed point gravity has a similar energy-dependence for $|s|/M_*^2\ll 1$. For $s\to M_*^2$, effective theory is no longer applicable and the energy dependence differs between the approaches. Effective theory predicts a weaker decay for large $s$ after analytical continuation as it is not sensitive to the fixed point behavior of the KK gravitons.  \medskip

To conclude, we obtain a good approximation for the amplitude  \eq{Cgeneral} within a principal value integration via the euclidean matching \eq{matchs+m}. The amplitude decays $\propto s^{-2}$ for large $s$ and  quantitatively becomes equivalent to  analytical continuation in $s$, \eq{Cs}.  The significance of the $s^{-2}$ scaling is that a necessary condition for perturbative unitarity is that the cross-section be bounded by $1/s$.  Examining the full $s$-channel amplitude we see that $C(s)\sim s^{-2}$ produces exactly this behavior at the level of $\sigma$. 
Hence, the amplitude decays asymptotically as required by perturbative unitarity for all physically motivated matchings. For $\Lambda_T/M_*$ up to the order of a few, the $s$-channel amplitude is unitary. In turn, for large $\Lambda_T/M_*\gg 1$, our results  fall back on those from  standard effective theory where perturbative unitarity is violated for center-of-mass energies approaching the fundamental Planck scale. Hence, the high-energy fixed point improves $s$-channel unitarity of the single graviton amplitude.

\subsection{Graviton width}\label{width}
\begin{figure}[t]
\includegraphics[scale=.9]{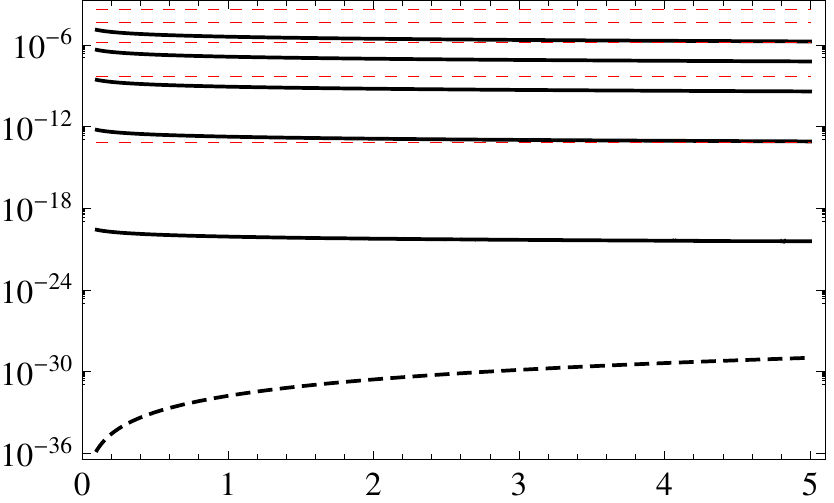}
\put(-55,-7){$\mkk\:(\tev)$}
\put(-100,35){$\Gamma$}
\put(-120,72){$\deltakk(n =2)$}
\caption[]{KK mass splitting $\deltakk$ (solid black) and the KK width
  $\Gamma$ (dashed black) as a function of $\mkk$, using
  $\mstar=5$~TeV.  The red dashed curves indicate the spacing
  $1/R$. Starting from the bottom counting upwards we have $n = 2-6$.}
\label{fig:widthdelta}
\end{figure}
In order to justify the virtuality matching, we discuss the narrow-width assumption 
for KK gravitons.  This aspect is independent of the UV sector, and also applies within 
standard effective theory. Using a Breit-Wigner propagator in the KK sum means we ignore 
interference between KK states, as the standard resummation ignores
mixing between different KK states.  In terms of the KK mass splitting 
this requires
\begin{equation}
\deltakk \gg \Gamma, 
\label{eq:funrelationwidth}
\end{equation}
for $\mkk \sim \sqrt{s}$ where $\deltakk$ is the spacing between consecutive
KK modes.  It is simple to compute the lowest order graviton decay width to
Standard Model particles, as it appears in \eq{eq:funrelationwidth}.  
This decay is suppressed by $1/\mpl^2$ since it occurs only on the brane. 
A calculation based on the Feynman rules given in Ref.~\cite{coll_tao} yields
\begin{equation}
\Gamma(G \to \text{SM}) = \frac{293}{960\pi} \; \frac{\mkk^3}{\mpl^2} \; .
\end{equation} 
We may also consider on-shell decays of a single
massive graviton to two other (less) massive gravitons.  This
contribution would produce a much larger decay width, suppressed
only by factors of $1/\mstar$.  However, extra-dimensional momentum conservation
persists as KK number conservation in the compactified theory
(also the basis for dark matter in UED models~\cite{ued}).  The occupation
vectors of all gravitons participating in an interaction must sum to zero at any
vertex.  When combined with energy momentum conservation
$|\vec{n}_1|^2> |\vec{n}_2|^2+|\vec{n}_3|^2$, the on-shell decay
amplitude vanishes for all graviton self interactions.  In fact, the
next contribution to the decay width will come from higher orders in
the expanded action
\begin{alignat}{5}
    \Gamma\;\;&\sim\text{Im}\left[
   \parbox{14mm}{ 
   \begin{fmfgraph*}(50,50)    
       \fmfleft{i}
       \fmfright{o}
      \fmf{double,tension=3}{i,v1}
       \fmf{fermion,left,tension=0.4}{v1,v2,v1}
       \fmfdot{v1,v2}
   \end{fmfgraph*}}
   	\quad\quad+\quad
   \parbox{14mm}{ 
   \begin{fmfgraph*}(60,60)
       \fmfleft{i}
    \fmfright{o}
  \fmf{double,tension=5}{i,v1}
       \fmf{double,tension=5}{v2,o}
    \fmf{fermion,left,tension=0.4}{v1,v2,v1}
       \fmf{double}{v1,v2}
       \fmfdot{v1,v2}
    \end{fmfgraph*}}
	\quad\quad\quad+\quad
   \parbox{14mm}{
   \begin{fmfgraph*}(50,50)    
       \fmfleft{i}
       \fmfright{o}
       \fmf{double,tension=3}{i,v1}
       \fmf{double,tension=3}{v2,o}
       \fmf{double,left,tension=0.4}{v1,v2,v1}
       \fmfdot{v1,v2}
   \end{fmfgraph*}}
   \quad + \; \cdots \right]
\notag \\
\notag \\
&\sim\quad \:\:\left[\quad\frac{\mkk^3}{\mpl^2}
\hspace{10.5mm}+\hspace{6.0mm}\frac{\mkk^3}{\mpl^2} 
\frac{\mkk^{n+2}}{\mstar^{n+2}}
\hspace{5.0mm}+\hspace{10mm} 0 \hspace{9.4mm} + \; \cdots \right],
\label{eg:onfeyndia}	
\end{alignat}
based on power counting. 
We emphasize again the null contribution due to
self-interactions.  The second diagram includes a sum over the
intermediate graviton as KK number is not conserved due to the
matter coupling. Indeed, below $\mstar$ the graviton width is
tiny, but at this scale the resummation quickly breaks down and the width
in general exceeds the spacing.  Therefore, we have little reason to
extend our classical results in the pole region above the scale $\mstar$.  
We emphasize that the gravitons contributing in this case 
probe IR physics.  A rephrasing of this physical picture
is that multi-graviton exchange such as the second diagram 
in \eq{eg:onfeyndia}  becomes relevant at the scale $\mstar$. It is 
still possible to use perturbation theory in the pole region above 
$\mtrans$ as long as we have a hierarchy $\mtrans<\mstar$. \medskip

The KK mass spacing is often estimated using $\deltakk =
1/R$~\cite{tasi_add}. We improve this estimate  to assure 
that the narrow-width condition in \eq{eq:funrelationwidth} does not break
down at scales significantly below $\mstar$. To that end, we
consider a state $\vec{n}_i$ where $\vec{n}_i^2 \equiv N$ is some
integer.  In general, there will exist another state $\vec{n}_{i-1}$
such that $\vec{n}_{i-1}^{2}=N-1$.  If there is no state satisfying
this relation then the spacing will be larger than our estimate,
leaving intact \eq{eq:funrelationwidth}.  The integer $N$ fixes
$\mkk = \sqrt{N}/R$ so that for two non-degenerate KK modes 
the spacing is at least
\begin{alignat}{5}
\deltakk&\ge \frac{\sqrt{N}-\sqrt{N-1}}{R} 
=  \left(\frac{(2\pi)^n  
	\mstar^{n +2}}
	{\mpl^2}\right)^{1/(2n) }\left[\sqrt{\mkk}	
	-\sqrt{\mkk-\left(\frac{(2\pi)^n  \mstar^{n +2}}
	{\mpl^2}\right)^{1/n}}\;\;
	\right] .
\label{eq:delspacej}
\end{alignat} 
We note that \eq{eq:delspacej} is sensitive to the
equality of the radii of the extra dimensions.  For the case of
unequal radii, the lower bound may be set by choosing the largest
radius.
In Fig.~\ref{fig:widthdelta} we see that this lower bound for the
spacing is much larger than the lowest order graviton width for any number of extra
dimension $n\ge 2$ and for all energy scales relevant to LHC physics.
For example, for $n=2$ and $\mstar=5\,\tev$ we  expect the lines to cross only
around $\mkk \sim 1000$~TeV, far above the scale at which the leading 
order approximation width breaks down. For the LHC observables computed 
in the following, \eq{eq:delspacej} confirms the hierarchy between 
the tiny KK width and the level spacing.

\section{Virtual gravitons at the LHC}
\label{sec:lhc}

The LHC has several on-going searches for extra-dimensions (for a recent example see \cite{CMS_bh}).  These include real graviton
emission, black-hole production and virtual graviton exchange.  Real emission searches 
are generally well-defined, and have little sensitivity to Planck scale physics at sub-Planckian collider energies.  
Similarly, black-hole production only becomes relevant at center-of-mass 
energies above the Planck scale. An important issue in understanding black hole production is the question of a remnant, {\sl i.e.} a minimum black hole mass which is predicted by fixed point gravity~\cite{Falls:2010he} as well as by string theory. Virtual graviton exchange probes the quantum gravity effects in the ultraviolet range of the Kaluza-Klein spectrum in a well defined field-theory environment. This makes this production process the prime candidate to test different descriptions of quantum gravity in low-scale gravity models. Hence, we apply our findings from the previous sections to gravitational Drell-Yan production at the LHC.

\subsection{Kinematics and cuts}
The generic prediction for virtual graviton exchange at the LHC is an
enhancement of rates at increasing invariant masses. This does not include
a mass peak or any kind of side-bin analysis, which means we must test 
the Standard-Model-only hypothesis as well as the 
Standard-Model-plus-graviton hypothesis on the kinematic distributions of 
the Drell-Yan process~\cite{grw,coll_tao}.  This signature may be 
further refined through the use of angular correlation techniques~\cite{ang_corr}.   Our
reference channel is $pp\to \mu^+\mu^-$ \cite{CMS_dimu} at $\sqrt{s}=14$~TeV but other
channels will show similar behavior. 
For the di-muon final state one
advantage from the theory point of view is that we need only consider
$s$-channel diagrams. \medskip

Based on the discussion in Sect.~\ref{sec:highC}, we evaluate the 
relevant amplitudes using analytic continuation, which sets a lower bound.
Moreover, since the SM background can originate only
from a $q\bar{q}$ initial state we have an automatic signal-over-background 
advantage due to probing the gluons in the initial state.
Therefore, the collider energy plays an obvious role affecting the discovery reach in two ways. First, for valence quarks in the initial states it determines the upper limit of the Kaluza-Klein tower we can probe. Second, because of the rapid drop of the gluon parton density in the proton the lower end of the Kaluza-Klein tower will be enhanced over the background significantly once the collider energy increases. However, unlike for black hole production the luminosity is the main limiting factor because of the relatively small and perturbative gravitational coupling. An optimized strategy for the discovery of virtual gravitons at the LHC given an energy and a luminosity is beyond the scope of this paper. Our theoretical considerations, on the other hand, will strongly impact such a study.
As model parameter we consider 2, 3 and 6 extra dimensions
and fix the fundamental Planck scale to $\mstar=5$~TeV.   
The experimental cuts which we apply throughout the section to take into account detector features are 
\begin{equation}
p_{T, \mu} > 50\gev \qquad \qquad \qquad 
|\eta_\mu| < 2.5 
\label{eq:cuts}
\end{equation}
The detection efficiency for muons we assume to be 100\%, including triggering, and their excellent energy resolution is perfect compared with the distributions we are going to look at.
In addition, whenever we compare signal and background rates we 
require $m_{\mu \mu}> \mstar/3$ to enhance the graviton signal over the Standard Model backgrounds.
Combined with the cuts in (\ref{eq:cuts}) this suppresses the highly IR peaked photon and $Z$ backgrounds which generically fall off with the partonic center-of-mass energy fast.  
As said above, the idea of this section is not to present 
a comprehensive LHC analysis,  but to test our results for the behavior
of the graviton KK integral at the level of LHC signals.

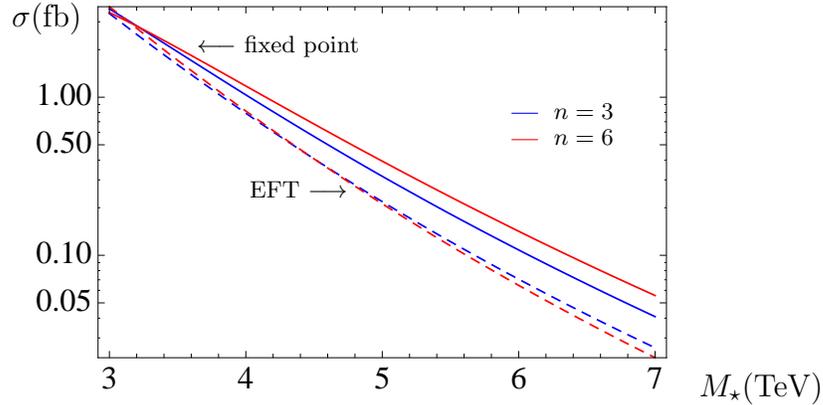
\begin{figure}[t]
\includegraphics[scale=1]{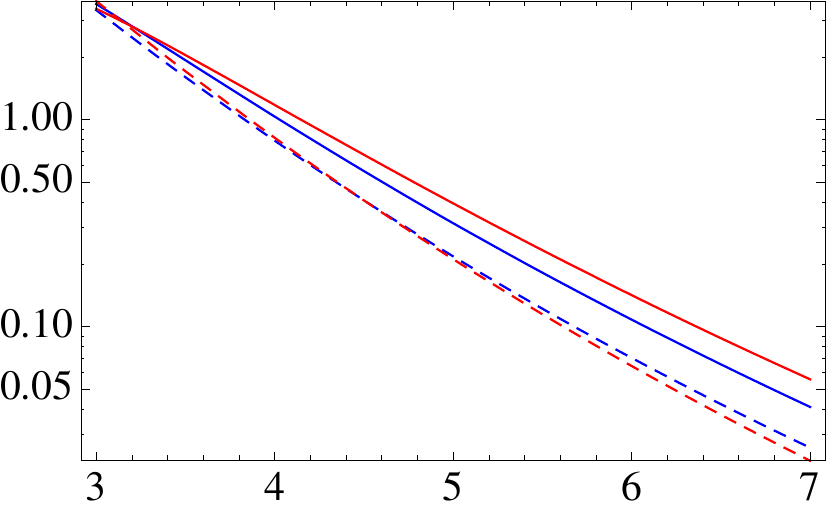} 
\put(-250,140){\large $\sigma(\text{fb})$ }
\put(10,0){\large $\mstar(\tev)$}
\put(-61,107){\color{blue}\line(2,0){10}}
\put(-45,105){$n=3$}
\put(-61,97){\color{red}\line(2,0){10}}
\put(-45,95){$n=6$}
\put(-180,130){$\longleftarrow$ fixed point}
\put(-160,75){EFT $\longrightarrow$ }
\caption{LHC total cross-section versus the fundamental scale of gravity 
$\mstar$  for events with $\sqrt{\hat{s}}<\mstar$ for $n=3$ (blue) and $n=6$ 
(red) extra dimensions. Solid lines: asymptotically safe gravity. Dashed lines: effective theory.}
\label{fig:lhce}
\end{figure}

\subsection{LHC signatures of the fixed point}

We first compare LHC predictions from fixed point gravity with those from effective theory.  Additionally we will evaluate the sensitivity of fixed point gravity signatures to the cross-over scheme, see Sect. \ref{anomalous} and Sect. \ref{scale_iden}.
The relevant amplitude is given by \eq{Cgeneral}. We compare the total cross section from fixed point gravity using $Z^{-1}$ from the quenched approximation \eq{quench} with the predictions from effective theory with UV cutoffs at $m=\Lambda_T$ and at $\Lambda_s=M_*$, which we implement as
$Z_{\rm eff}^{-1}=\theta(\mtrans-\mkk)\,\theta(\mstar-\sqrt{s})$.
The result  is displayed in  Fig.~\ref{fig:lhce} for $M_*=5$~TeV and $\Lambda_T=M_*$. Dashed lines denote the result from effective theory,
full lines stand for fixed point gravity.
The difference between effective theory and fixed point gravity (solid versus dashed curves) is negligible at low scales $s\ll M^2_*$, provided the UV cutoff in effective theory is identified with the cross-over scale, $\Lambda=\Lambda_T$. With increasing energy $\sqrt{s}\approx M_*$, the difference becomes more pronounced, indicating the limit of validity for an effective theory description.  
It turns our that with increasing $n$ the signal becomes more sensitive to the UV modes.  This is exactly what is expected based on the discussion in Sect.~\ref{KKsector}.\medskip

\begin{center}
\vskip-.4cm
\begin{table}
\begin{tabular}{|c|rr|rr|}  
  \hline
   $\sigma [{\rm fb}]$  & \multicolumn{2}{c|}{$n=3$} & \multicolumn{2}{c|}{$n=6$}\\[1mm]
   $M_*$[TeV]             & $5$ & $8$  & $5$ & $8$ 
\\
  \hline
   $a)$     &\; 0.317  & \;0.015 & \;0.393  & \;0.020 \\
   $b)$     &\; 0.320  & \;0.015 & \;0.394  & \;0.020 \\        
   $c)$     &\; 0.215  & \;0.009 & \;0.209  & \;0.007 \\
$d)$          &\; 0.182  & \;0.009 & \;0.190  & \;0.007 \\
\hline
\end{tabular}  
\caption{\label{T1}
Comparison of total cross-sections for di-muon production via virtual graviton exchange at the LHC in different approximations.  
a) fixed point gravity with $\Lambda_T=M_*$, and an energy cutoff at $\Lambda_s=M_*$
b) fixed point gravity with $\Lambda_T=M_*$ (no energy cutoff)
c)  effective theory with UV cut-offs at $\Lambda_{\rm kk}=\Lambda_s=M_*$,  
d) form factor approximation with  cutoff at $\Lambda_{\rm kk}=M_*$  (see text). 
}
\end{table}
\end{center}

In Tab.~\ref{T1} we compare the production cross-sections in various 
approximations for $\mstar=\mtrans=5\,\tev$.  For the UV 
contributions arising from KK modes, comparing fixed point gravity a,b) with effective theory c), we find  
an increase in the total rate.  
On the other hand, including  the UV events in $\sqrt{s}$ by comparing $a)$ with $b)$ we find only 
a modest increase.  Evidently, the $1/s^2$ behavior for large energies makes these contributions sub-leading.
The form factor approximation $d)$ comes out below the effective theory estimate $c)$, introducing a modest extra reduction of overall rates with energy at below-Planckian energies.
Provided the cross-over scale takes the larger value $\mstar= 8\,\tev$, 
we find that the LHC is only very weakly  sensitive to  the UV KK modes, leaving no difference between $a)$ and $b)$, and $c)$ and $d)$, respectively. Still, the difference between $a), b)$ and $c), d)$ persists.
\medskip

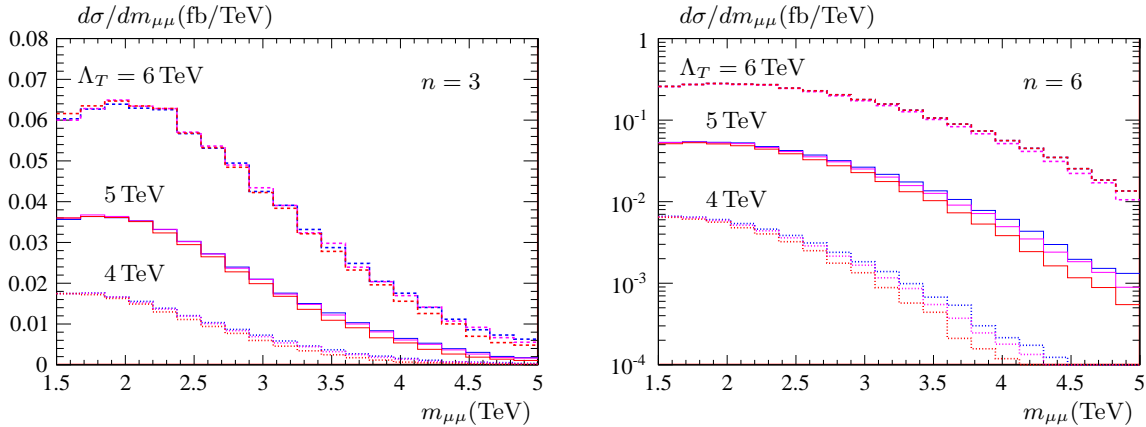
\begin{figure}[t]
$
\begin{array}{cc}
\includegraphics[scale=.4]{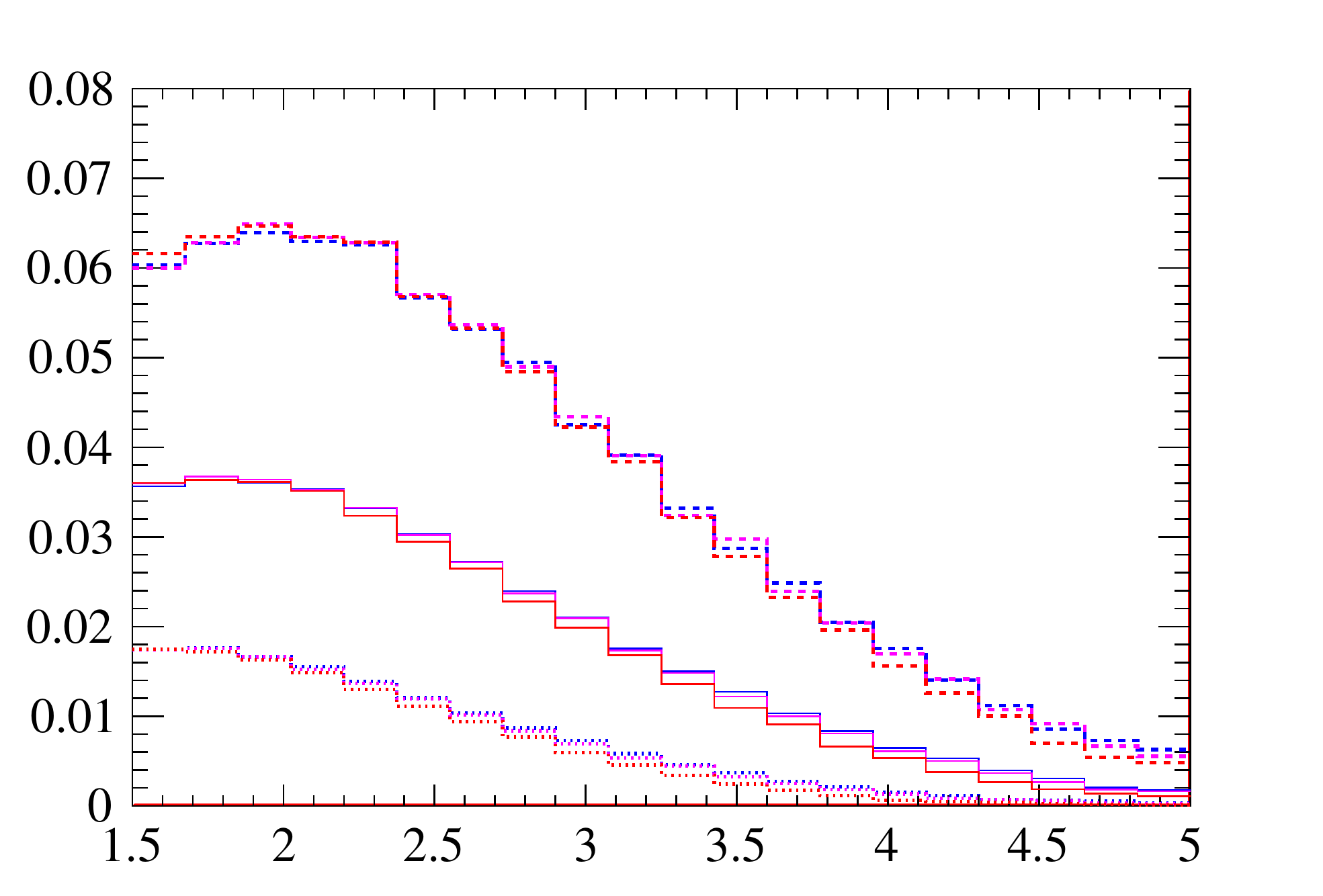} 
\put(-200,145){
${d\sigma}/{dm_{\mu\mu}}{
(\text{fb}/\tev})$ }
\put(-200,122){ $\mtrans=6\,\tev$}
\put(-190,77){ $
5\,\tev$}
\put(-190,47){ $
4\,\tev$}
\put(-70,-5){
$m_{\mu\mu}(\tev)$ }
\put(-70,120){
$n=3$ }
\includegraphics[scale=.4]{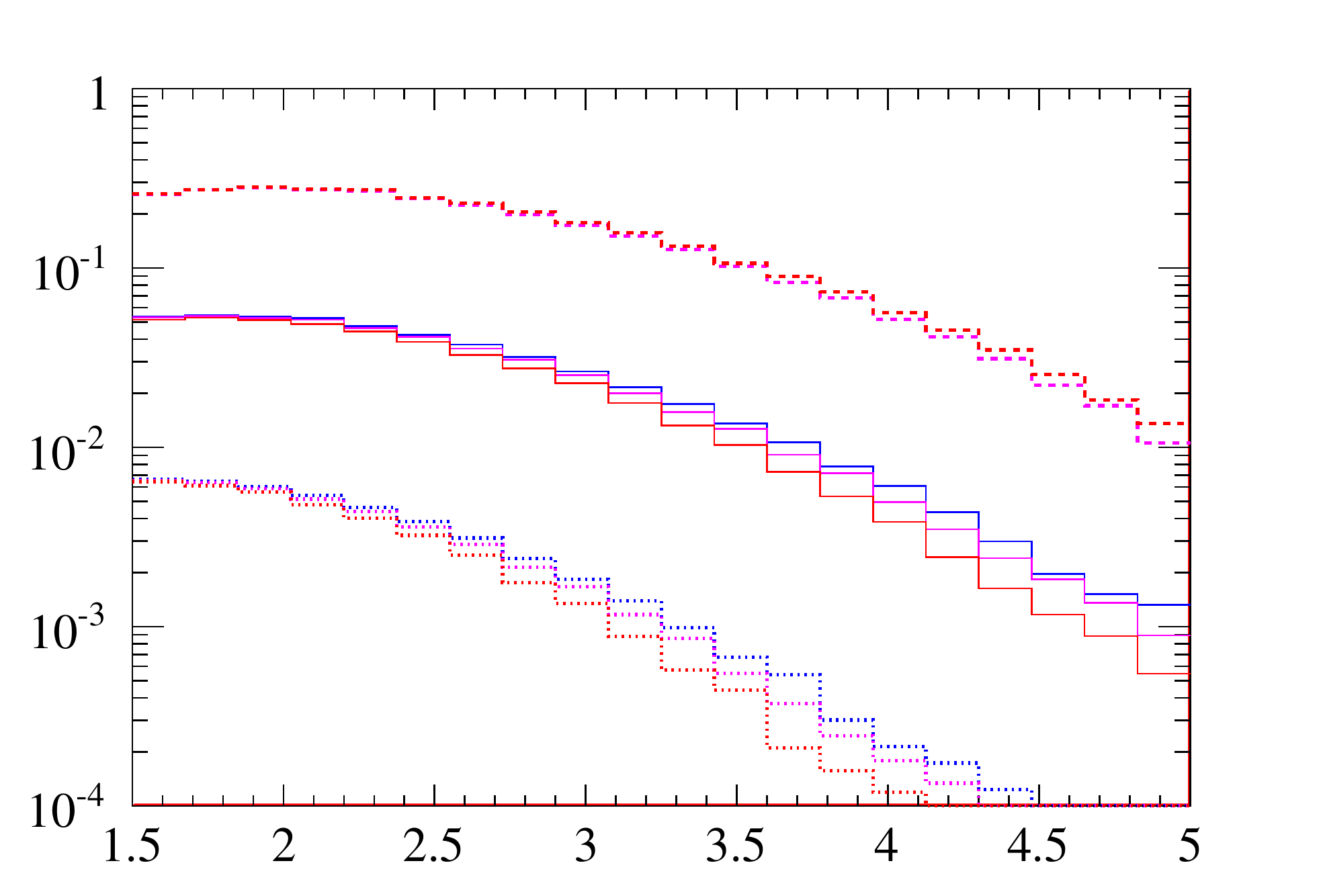} 
\put(-200,145){
${d\sigma}/{dm_{\mu\mu}}{
(\text{fb}/\tev})$ }
\put(-70,-5){
$m_{\mu\mu}(\tev)$ }
\put(-70,120){
$n=6$ }
\put(-200,125){ $\mtrans=6\,\tev$}
\put(-190,105){ $
5\,\tev$}
\put(-190,75){ $
4\,\tev$} 
\end{array}
$
\caption{Signal distribution for di-muon production via virtual graviton exchange 
at the LHC with $M_*=5$~TeV and $n=3$ $(6)$ extra dimensions show left (right). Comparison of several RG schemes and transition scales $\Lambda_T$: quenched (red lines), linear (blue), and quadratic (magenta) approximation with $\Lambda_T/M_*=1.2 $ (dotted lines), $\Lambda_T/M_*=1$ (full),  and $\Lambda_T/M_*=0.8$ (dashed). With increasing $\Lambda_T/M_*$, the di-muon production rate increases. The scheme dependence is moderate.} 
\label{fig:compareM31}
\end{figure}

Next we compare the results from fixed point gravity using the quenched \eq{quench}, linear \eq{linear} and quadratic \eq{quadratic} forms of the RG running, at fixed transition scale $\Lambda_T$. The variation of results under changes in the RG running allow an estimate for the `RG scheme dependence' in the present set-up. The weak scheme dependence of the $s$-dependent amplitudes was already demonstrated in Sect.~\ref{Planck}, see also Fig. \ref{fig:Compare} .  We would like to verify this at the level of LHC cross-sections.
In Fig. \ref{fig:compareM31} we establish scheme independence for $n=3$ and $n=6$ extra-dimensions, respectively, for several values of the transition scale $\Lambda_T$.  
The variation is minute at the low-end of the spectrum but increases at higher energies as expected.  
In all cases the uncertainty due to the scheme is smaller than uncertainties coming from QCD effects or 
from parton densities, conservatively estimated at $\sim15 \%$ here.  \medskip

A second point illustrated in Fig.~\ref{fig:compareM31} is that
the total rates depend strongly on the transition scale $\Lambda_T$.  This is especially true for higher $n$, as has been pointed out in Sect.~\ref{KKsector}, see Fig.~\ref{fig:NDependence}.  The reason for this is that the suppression of amplitudes due to the gravitational fixed point sets in at about $\Lambda_T$, leading to larger rates for larger $\Lambda_T$. However, the $s$-dependence in the signal is mostly independent of $\Lambda_T$. 
 In the next section we study the normalized distributions in order to obtain a handle on the signals.

\subsection{Phenomenology}

In this section, we discuss signatures of asymptotically safe gravity and large extra dimensions  at the LHC.
If not stated otherwise, we set $\mstar=\mtrans=5\,\tev$ and examine 
distributions with respect to $m_{\mu\mu}$, the invariant mass of the produced 
di-muon pair.  We also compare  our 
results, and in particular  the $m_{\mu\mu}$-dependence, with those from effective field 
theory \cite{grw,gps,Giudice:2003tu}, the form-factor approximation
 \cite{tom_joanne}, and the Standard Model background. \medskip

In Fig.~\ref{fig:compareM2},  we compute normalized and 
un-normalized differential cross sections for $M_*=5$~TeV and $\Lambda_T=M_*$ for
 $n=2,3,6$ extra dimensions.  Again, we contrast fixed point gravity results with those from effective theory.
Comparing the full $\sqrt{s}$-dependence from a) asymptotically safe gravity with b) effective theory (full versus dashed colored lines), we note that they essentially agree for $n=2$. The reason for this is that the perturbative amplitude is only logarithmically divergent. For $n=3$ and $6$ the decay of the differential cross section with invariant mass becomes more pronounced within fixed point gravity. Here, the difference is due to the UV divergence of the perturbative amplitude. Note in particular that with increasing number of extra dimensions $n$,  the difference between a) and b) grows. 
Turning to the low-$s$ approaches, we note that the differential cross section decays more slowly within the form-factor approximation c) (full black lines) for all $n$. This comes about because a dynamical suppression with energy sets in only close to $\sqrt{s}\approx\Lambda_T$, unlike a). Also, the form-factor approximation c) deviates from the effective theory approximation (with $s=0$) d) only close to $\Lambda_T$.    
Overall, we see that fixed point gravity falls off most quickly with increasing energy $\sqrt{s}$.  
Technically, this results from the powers of $s$ in the denominator of the RG improved KK sum, not present in  the lowest dimensional operator from effective theory. 
  \medskip

\begin{figure}[t]
\includegraphics[scale=.5]{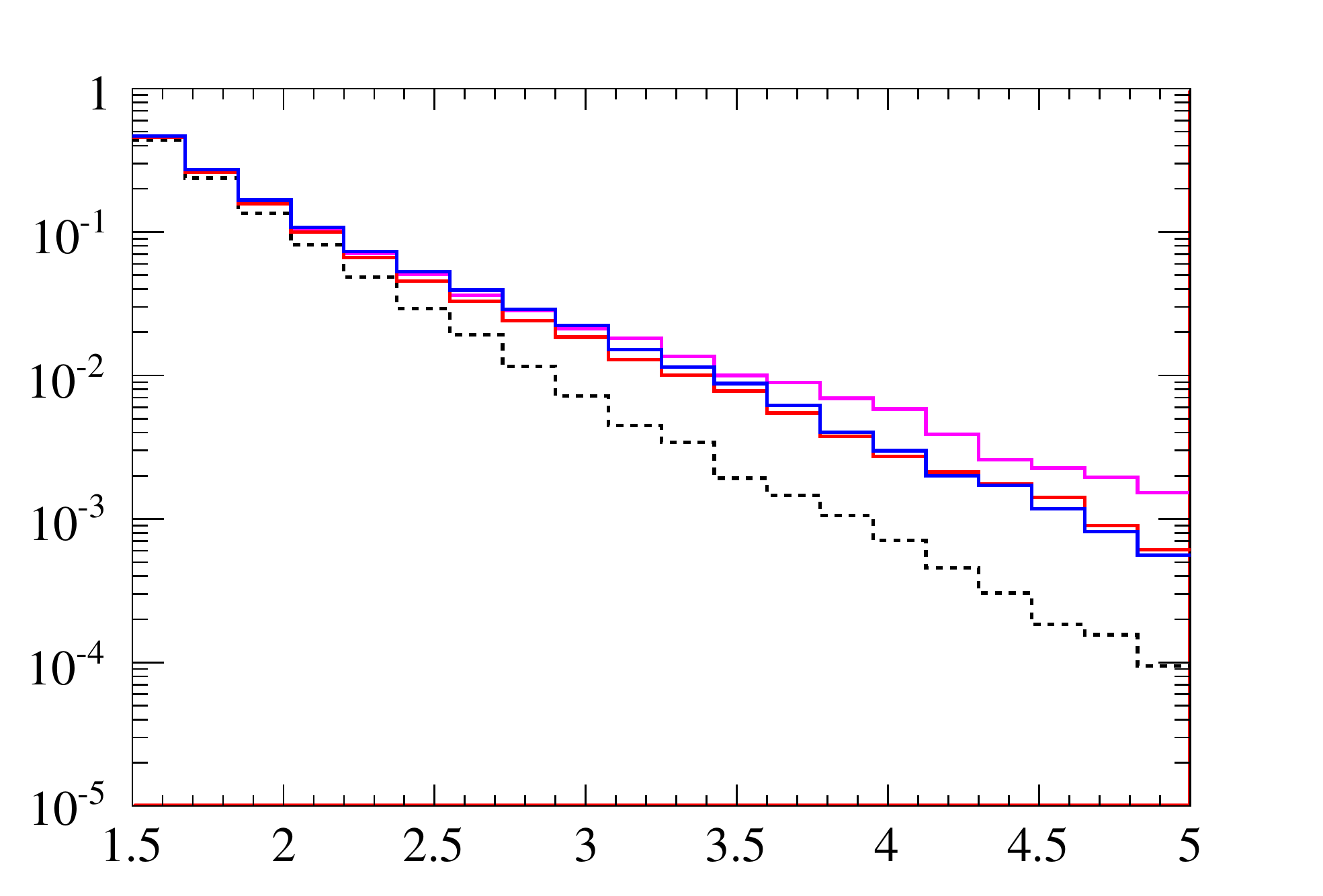} 
\put(-340,155){\large $\dfrac{d\sigma}{dm_{\mu\mu}}{\large (\text{fb}/\tev)}$ }
\put(-10,5){\large $m_{\mu\mu}({\rm TeV})$ }
\put(-150,60){Standard Model $\rightarrow$}
\linethickness{.3mm}
\put(-121,157){\color{magenta}\line(2,0){10}}
\put(-105,155){$n=2$}
\put(-121,147){\color{blue}\line(2,0){10}}
\put(-105,145){$n=3$}
\put(-121,137){\color{red}\line(2,0){10}}
\put(-105,135){$n=6$}
\put(-100,110){$\leftarrow$ FP}
\caption{Differential cross-sections (in fb/TeV) for di-muon production within fixed point  gravity (FP) for $n=2$ (magenta), $n=3$  (blue) and $n=6$ (red) extra dimensions, in comparison with Standard Model background (black dashed line). }
\label{fig:SM}
\end{figure}
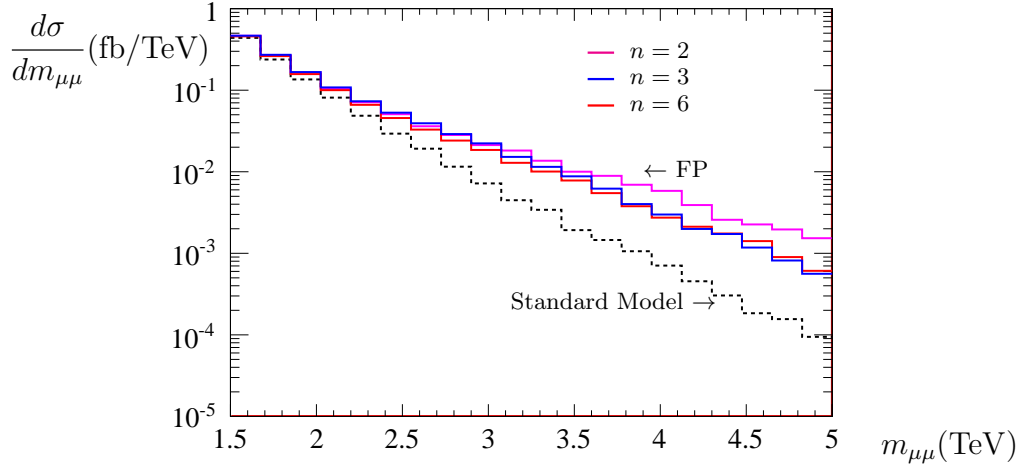

In Fig.~\ref{fig:SM} we present a comparison of our data including the 
SM background.  For larger values of $m_{\mu\mu}\sim \mstar/2$ signal to background
$S/B$ becomes of order 1.  Note that the cross section for $n=2, 3$ and $n=6$ are very close to each other. 
This near-degeneracy  is a consequence of our choice $\Lambda_T=M_*$, which is lifted as soon as 
$\Lambda_T\neq M_*$, see Fig.~\ref{fig:NDependence}.  
In view of perturbative unitarity, our calculations 
in~(\ref{Cdecay}) and the behavior in Fig.~\ref{fig:compareM2}
indicate a clear improvement over effective theory. \medskip

Phenomenologically, for the LHC we now know that the on-set of fixed point scaling  should be distinguishable form eg.~effective field theory descriptions. Turning this argument around, the ad hoc assumption of an effective field theory is not well suited to analyze LHC data which should be expected to strongly depend on ultraviolet effects. This is even more obvious for string theory signatures~\cite{lhc_strings} with their peak structure dictated by the Veneziano amplitude. But also the two existing descriptions of quantum gravity at the LHC, fixed point gravity and string theory could not look more different in their predictions for the $m_{\mu \mu}$ distribution. Fig.~\ref{fig:SM} provides a realistic signature for asymptotic safety. Moreover, the differential cross section as a function of $m_{\mu\mu}$ can be used to distinguish  our effect uniquely and study in some detail the transition from the classical to the quantum gravity regime.

\begin{center}
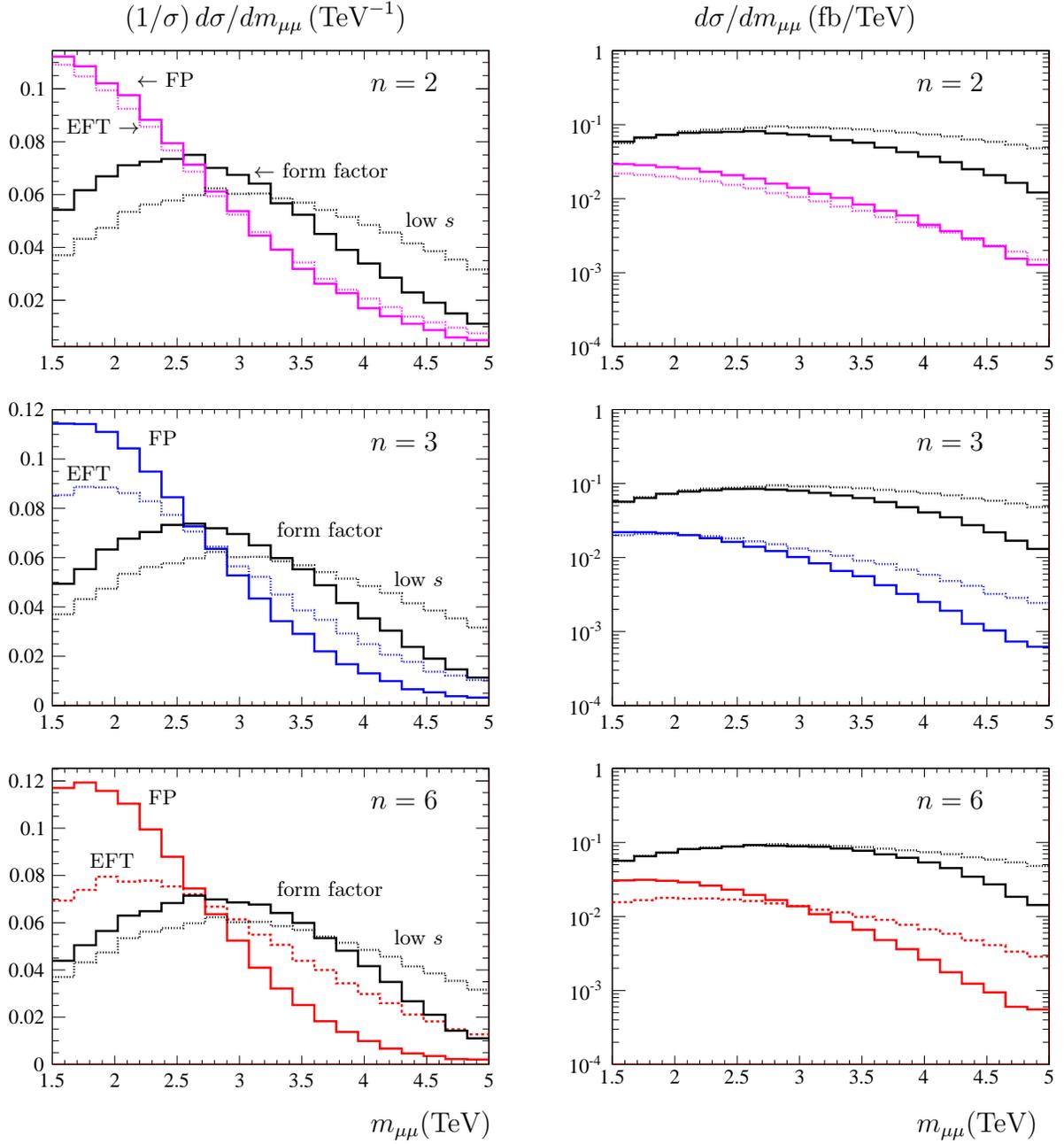
\begin{figure}[ht]
$
\begin{array}{cc}
\includegraphics[scale=.41]{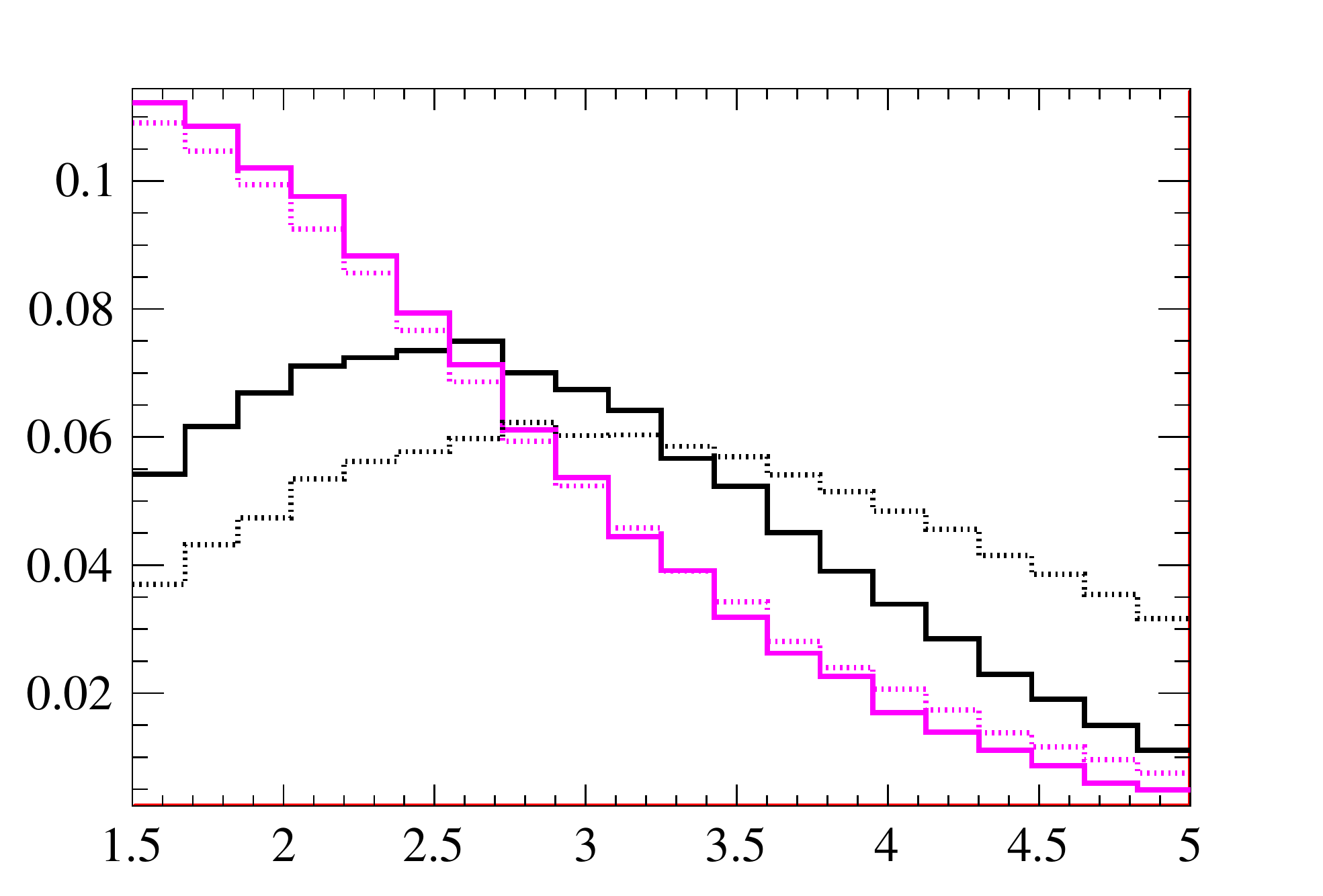} &
\put(-185,153){\large ${(1/\sigma)\,d \sigma}/{dm_{\mu\mu}}\,{(\tev^{-1})}$ }
\put(-80,125){\large $n=2$ }
\put(-180,127){$\leftarrow$ FP}
\put(-210,107){EFT $\rightarrow$}
\put(-130,88){$\leftarrow$ form factor}
\put(-65,67){low $s$}
\includegraphics[scale=.41]{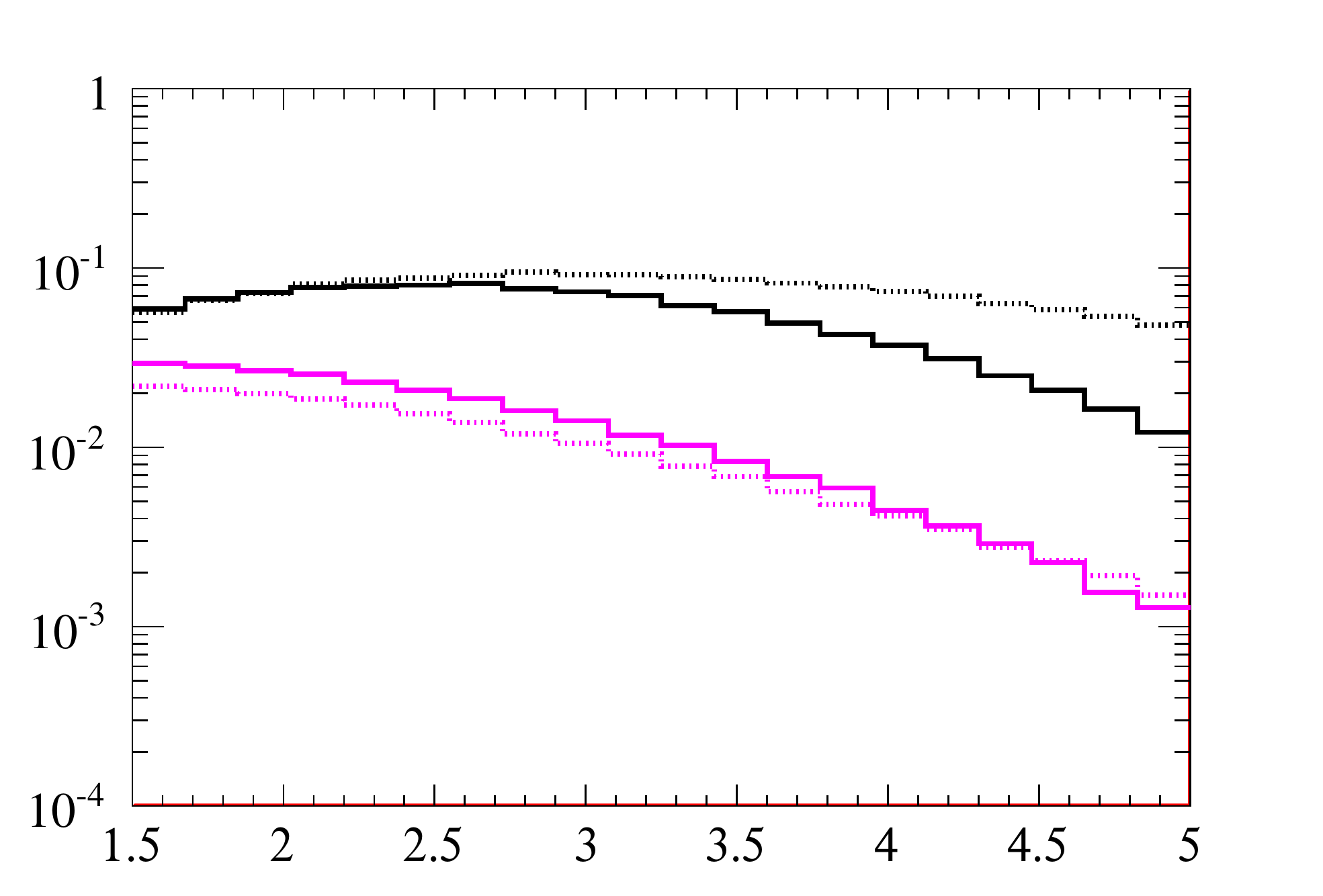}
\put(-175,153){\large ${d\sigma}/{dm_{\mu\mu}}\,{(\text{fb}/\tev)}$ }
\put(-80,125){\large $n=2$ }
\\[-2ex]
\includegraphics[scale=.41]{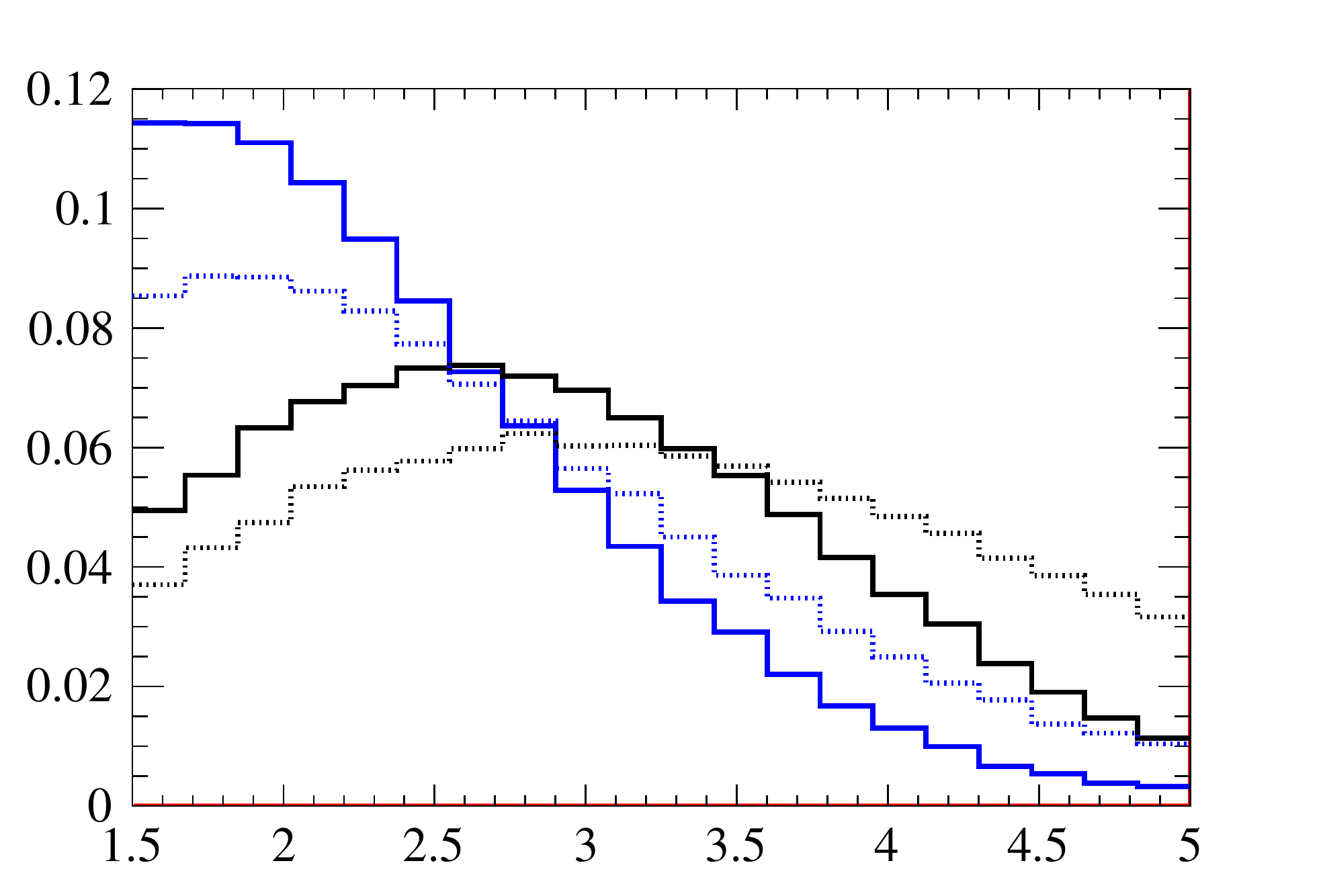} &
\put(-80,125){\large $n=3$ }
\put(-175,127){FP}
\put(-210,112){EFT}
\put(-120,88){form factor}
\put(-70,67){low $s$}
\includegraphics[scale=.41]{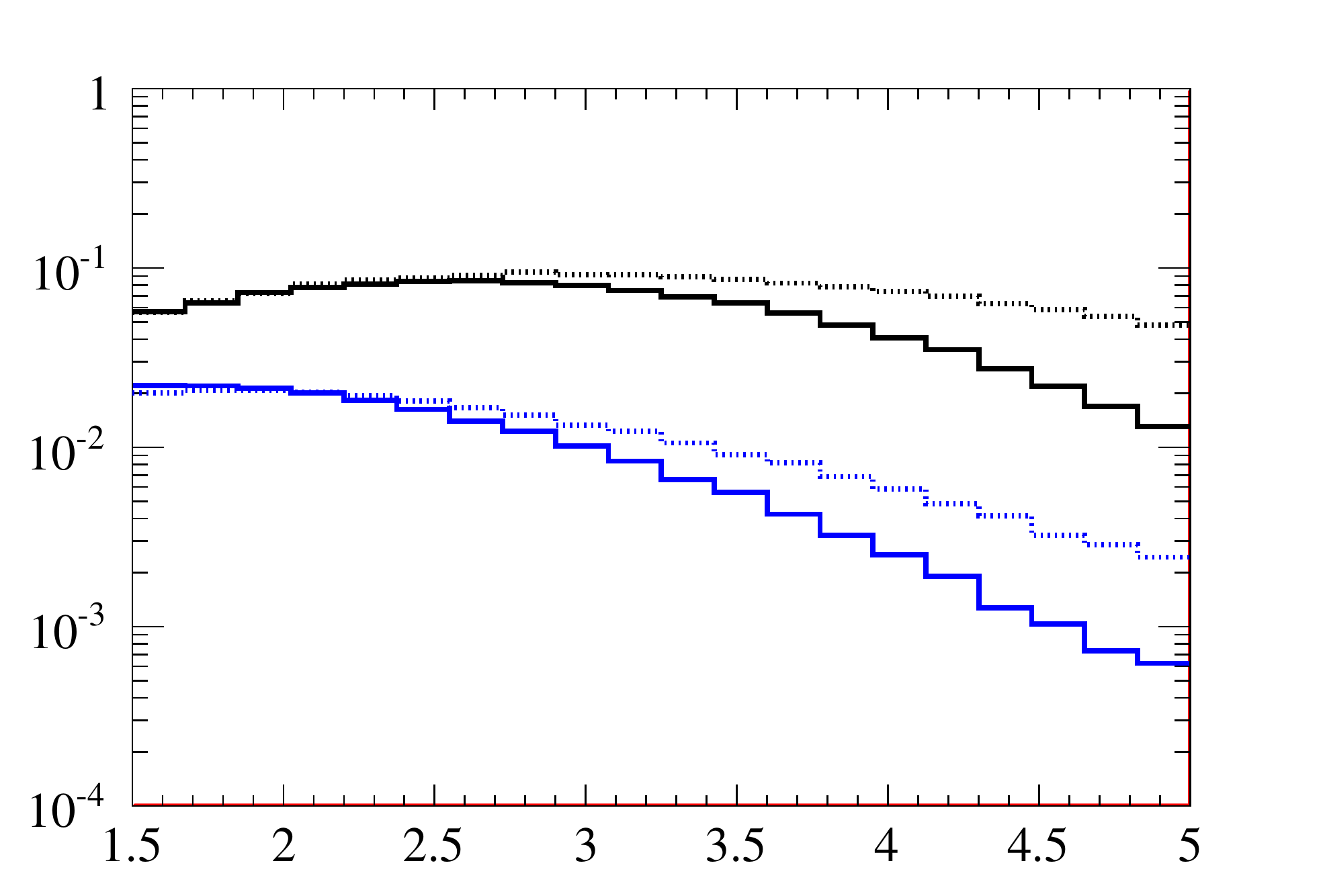}
\put(-80,125){\large $n=3$ }
\\[-2ex]
\includegraphics[scale=.41]{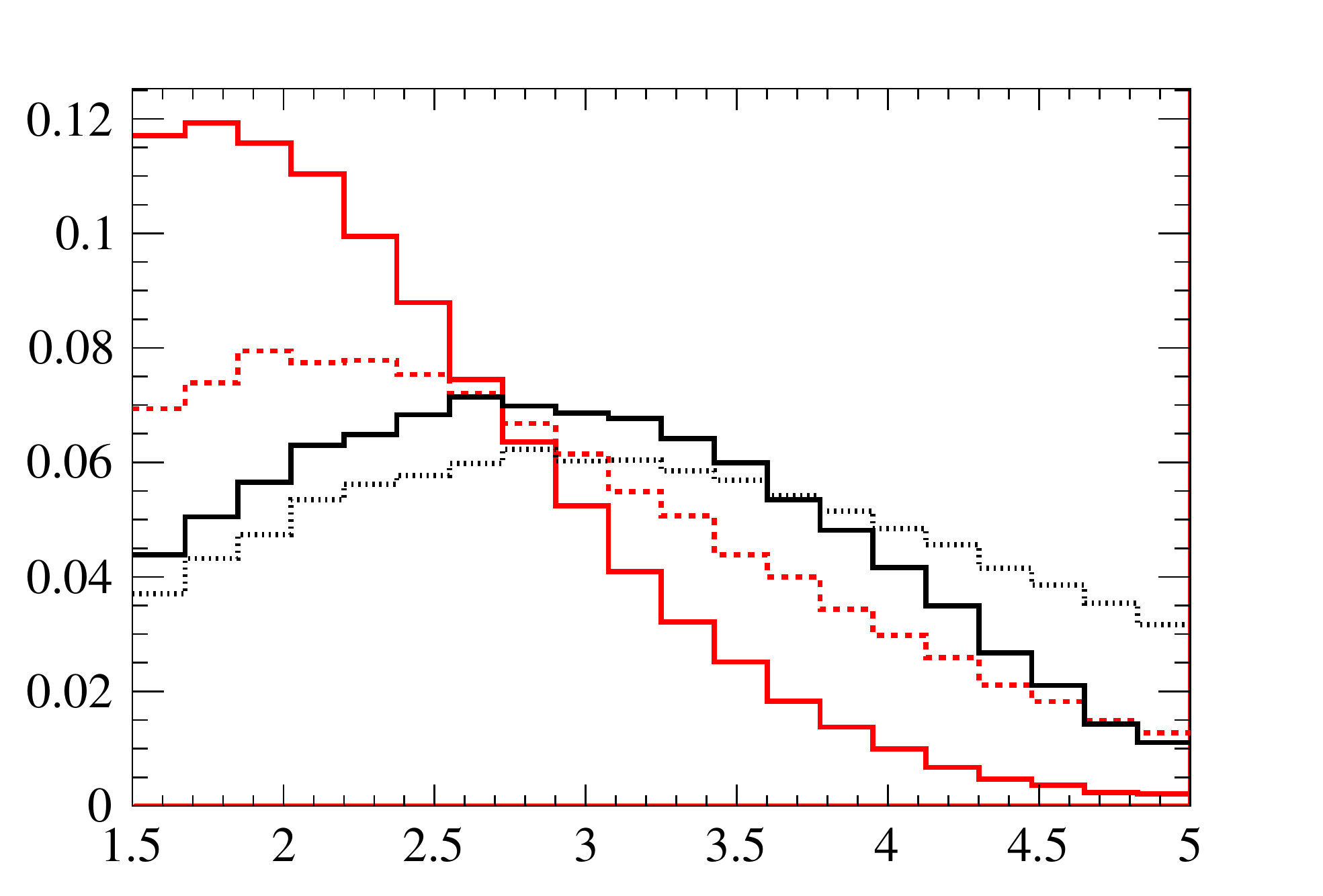} &
\put(-80,-13){\large $m_{\mu\mu}$(TeV) }
\put(-80,125){\large $n=6$ }
\put(-175,127){FP}
\put(-200,100){EFT}
\put(-120,88){form factor}
\put(-70,67){low $s$}
\includegraphics[scale=.41]{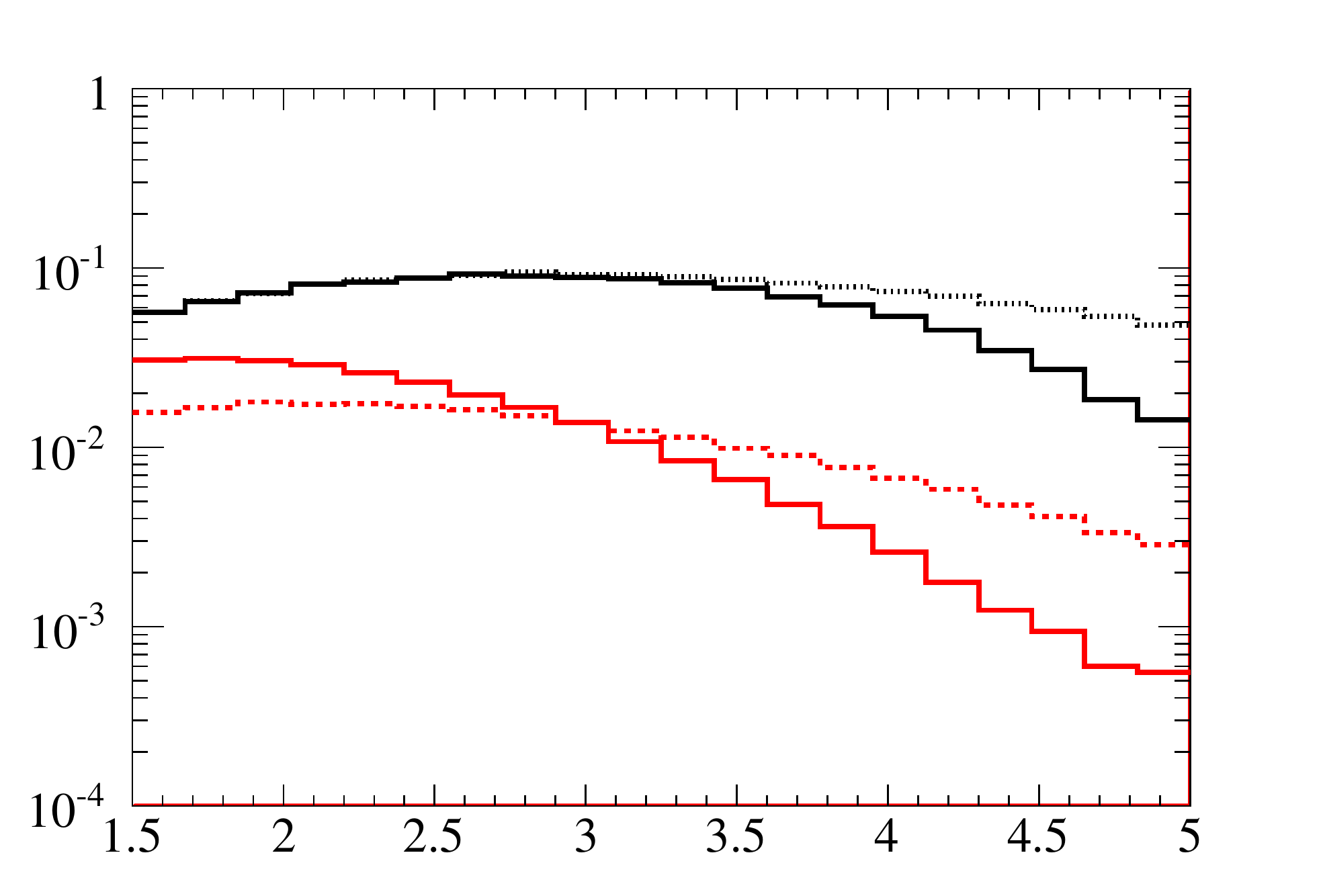}
\put(-80,125){\large $n=6$ }
\put(-80,-13){\large $m_{\mu\mu}$(TeV) }
\end{array}$
\caption{Normalized differential cross sections $(1/\sigma)\, d\sigma/dm_{\mu\mu}$ (left column) and un-normalized ones  $d \sigma/dm_{\mu\mu}$ (right column)  for gravitational di-muon production 
at the LHC for $n=2, 3$ and $6$ (top, middle, bottom row) with $M_*=5$~TeV and cross-over scale $\Lambda_T=5$~TeV.
a) fixed point gravity (FP) with  full $s$-dependence  (full magenta/blue/red lines); b) fixed point gravity in the form factor approximation 
 with UV cutoff at $m=\Lambda_T$ (full black lines); c)  effective theory (EFT) 
 with UV cutoff at $\Lambda_T$ (dotted magenta/blue/red lines); d)  effective theory in the $s=0$ approximation (low $s$) and UV cutoff at $\Lambda_T$ (dotted grey lines). 
} 
\label{fig:compareM2}
\end{figure}
\end{center}

\section{Discussion}\label{discussion}

We studied gravitational di-lepton production at the LHC, provided  gravity displays a non-trivial fixed point at high energies \cite{Litim:2003vp,Fischer:2006fz}. Previously,  it was shown that the RG improved single graviton amplitude becomes finite \cite{lp}. Furthermore, with increasing $n$ the UV Kaluza-Klein gravitons dominate over the IR Kaluza-Klein gravitons, leading to an increased production cross section over estimates from effective theory in the kinematical regime with $s\ll M_*^2$.  \medskip

Here, we have extended this study towards intermediate center-of-mass energies. We have worked out how the RG scale links with the kinematics and the Kaluza Klein mass and compared several scale identifications such as  \eq{matchm}, \eq{matchs+m}, \eq{matchs}, and \eq{abs}.  We have stressed the importance of incorporating the KK mass,  as done in  \eq{matchm}, \eq{matchs+m} and \eq{abs}, and evaluated the amplitude via analytic continuation in $s$. We found a suppression of the amplitude at Planckian energies (Fig.~\ref{fig:coefficientS}) compared to the low energy amplitude.
The extra-dimensional Kaluza-Klein gravitons act as messengers of the UV fixed point in the higher-dimensional theory, whose existence leads to a dynamical suppression of the single graviton amplitude on the brane. This phenomenon is dictated by the fixed point, and largely independent of the number of extra dimensions. Furthermore, with increasing $n$, the dynamical suppression sets in at lower scales.  This pattern is neither visible within effective theory nor within form factor approximations as these operate a cutoff which screens the fixed point dynamics of the KK gravitons.  Provided the dynamics of the KK modes are taken into account, differential cross section show a decrease with invariant mass, compatible with perturbative unitarity.  Furthermore,  the width of the transition regime becomes narrower with increasing $n$ \cite{Litim:2003vp,Fischer:2006fz}, thus increasing the sensitivity to the scale $\Lambda_T$.\medskip

Our results also have an interesting interpretation from a four-dimensional (brane) perspective. The effective single graviton amplitude $C_{\rm eff}(s)$ obtained {after} integrating-out the KK modes can be written in terms of an effective four-dimensional running gravitational coupling $C_{\rm eff}(s)\sim G_{\rm eff}(\sqrt{s})/s$. In the absence of extra dimensions, $G_{\rm eff}(\mu)$ reduces to Newton's constant $G_N$. In the presence of extra dimensions, the scale-dependence of $G_{\rm eff}(\mu)$ is induced by the extra-dimensional KK gravitons, and the gravitational fixed point in the higher dimensional theory. Interestingly, 
after integrating out the extra-dimensional gravitons, $G_{\rm eff}(\mu)$ behaves as $\propto 1/\mu^2$ for large $\mu$, which corresponds  to an effective four-dimensional fixed point in the dimensionally reduced theory. \medskip

The single graviton amplitude involves the on-shell production of massive KK modes. We evaluated the amplitude via a principal value prescription, and verified the narrow-width assumption. In general, within fixed point gravity the growth of the amplitude with energy is reduced compared to effective theory. It would be interesting to re-visit this approximation for KK masses reaching the fundamental Planck scale. We also neglected the production of real gravitons, which can leave the detector unobserved. Their presence can be detected indirectly, for example via missing energy signatures. \medskip

For large trans-Planckian energies the single graviton amplitude becomes sub-leading and multi-graviton effects will take over. One could speculate that the cross section should be dominated by the formation of gravitational bound states such as mini-black holes. Within asymptotically safe gravity, the production cross section for quantum black holes displays a threshold at a critical black hole mass $M_c$, and reaches a semi-classical limit for asymptotically large $\sqrt{s}\gg M_*$ \cite{Falls:2010he}. It would be interesting to see the on-set of quantum black hole production from multi-graviton scattering \cite{'tHooft:1987rb}. For recent developments along these lines within string quantum gravity, see \cite{Amati:2007ak,Marchesini:2008yh}. \medskip

We also compared our results with those from effective theory, where amplitudes depend on an unknown ultraviolet cutoff scale $\Lambda_{\rm kk}$. This scale is fixed once the fundamental theory for gravity is known. Formally, our results reduce to those from effective theory in the limit where the transition scale  $\Lambda_T$ becomes asymptotically large. In this case, the running of Newton's coupling is switched off. Also, the UV cutoff of the effective theory  can be expressed through the fundamental scale $\Lambda_T$ and the energy. This map shows that the domain of validity for effective theory varies parametrically with $n$, because the amplitude becomes sensitive to the UV completion already for center-of-mass energies below the fundamental Planck scale. \medskip

From an experimental point of view, it will be most important to determine the scale $\Lambda_T$.  We saw that the single graviton amplitude is sensitive to $\Lambda_T$, which sets the overall normalization of production cross sections. Interestingly, the amplitude is largely insensitive to the details of the cross-over.  Slight variations in the result are below other experimental and theoretical uncertainties.  Therefore, we expect that the qualitative change of the amplitude due to the fundamental cross-over scale $\Lambda_T$ are stable predictions of our scenario.  The details of the UV fixed point, such as the momentum dependence of the amplitude,  could in principle be studied by careful examination of the distributions with respect to the di-muon invariant mass. \medskip

\section{Conclusion}\label{conclusions}

Extra-dimensional models with a fundamental Planck scale as low as the electro-weak scale 
offer an excellent opportunity to test the quantization of gravity at colliders.
If gravity is asymptotically safe, gravitational interactions become weaker towards high energies.
The onset of asymptotic safety is characterized by the energy scale $\Lambda_T$  which marks the transition from Newtonian scaling at low energies to fixed point scaling at high energies.
We find that graviton-induced Drell-Yan spectra are well above Standard Model backgrounds with a significant sensitivity for measuring  $\Lambda_T$ and thus  conclude that Drell-Yan production has exciting prospects in probing quantum gravity at the LHC.

\subsection*{Acknowledgments}

EG is grateful to the Institut f\"ur Theoretische Physik at Heidelberg
University for their continuous hospitality. Moreover, EG and TP would like
to thank Jan Pawlowski for all the encouraging discussion and for his
theoretical insights.
DL thanks Gudrun Hiller for discussions and the Aspen Center for Physics for hospitality. 
This work was supported by the Science and Technology Research Council [grant number ST/G000573/1].
\bigskip

Note added: after finalizing this paper a study of LHC
bounds on virtual gravitons appeared which has some overlap with part
of the discussions presented above~\cite{note_added}.

\end{fmffile}


\end{document}